\documentclass[lettersize,journal]{IEEEtran}
\usepackage{amsmath,amsfonts}
\usepackage{algorithmic}
\usepackage{algorithm}
\usepackage{array}
\usepackage[caption=false,font=normalsize,labelfont=sf,textfont=sf]{subfig}
\usepackage{textcomp}
\usepackage{stfloats}
\usepackage{url}
\usepackage{verbatim}
\usepackage{graphicx}
\usepackage{cite}
\usepackage{lipsum}
\usepackage{booktabs}
\usepackage{gensymb}
\usepackage{xcolor}
\usepackage{hyperref}
\usepackage{amsmath,amsfonts,amssymb,amsthm,bbm}
\usepackage{caption}

\newcommand{\epsmethod}{\textit{AvatarLDP}}
\newcommand{\thetamethod}{\textit{AvatarRotation}}

\newtheoremstyle{theoremdd}% name of the style to be used
  {\topsep}% measure of space to leave above the theorem. E.g.: 3pt
  {\topsep}% measure of space to leave below the theorem. E.g.: 3pt
  {\itshape}% name of font to use in the body of the theorem
  {16pt}% measure of space to indent
  {\bfseries}% name of head font
  {\newline}% punctuation between head and body
  { }% space after theorem head; " " = normal interword space
  {\thmname{#1}\thmnumber{ #2}\textnormal{\thmnote{ (#3)}}}

\theoremstyle{theoremdd}

\newtheorem{theorem}{Theorem}
\newtheorem{definition}{Definition}

\newtheorem{corollary}{Corollary}

\def\editmode{1}

\newcommand{\todo}[1]{%
    \if\editmode1%
        \textcolor{blue}{\textbf{TODO: #1}}%
    \fi%
}

\newcommand{\note}[1]{%
    \if\editmode1%
        \textcolor{blue}{\textbf{NOTE: #1}}%
    \fi%
}

\begin{document}

\title{Towards Privacy-preserving Photorealistic Self-avatars in Mixed Reality}

\author{Ethan Wilson$^1$, Vincent Bindschaedler$^1$, Sophie J\"{o}rg$^2$, Sean Sheikholeslam$^1$, Kevin Butler$^1$, Eakta Jain$^1$%
\thanks{1. University of Florida, 2. University of Bamberg.}%
\thanks{Manuscript received X; revised Y.}}

% The paper headers
%\markboth{IEEE TRANSACTIONS ON VISUALIZATION AND COMPUTER GRAPHICS, VOL. X, NO. X}{Wilson \MakeLowercase{\textit{et al.}}: Towards Privacy-preserving Photorealistic Self-avatars in Mixed Reality}

% \IEEEpubid{0000--0000/00\$00.00~\copyright~2021 IEEE}
% Remember, if you use this you must call \IEEEpubidadjcol in the second
% column for its text to clear the IEEEpubid mark.

\maketitle

\begin{abstract}
Photorealistic 3D avatar generation has rapidly improved in recent years, and realistic avatars that match a user's true appearance are more feasible in Mixed Reality (MR) than ever before.  Yet, there are known risks to sharing one's likeness online, and photorealistic MR avatars could exacerbate these risks.  If user likenesses were to be shared broadly, there are risks for cyber abuse or targeted fraud based on user appearances.  We propose an alternate avatar rendering scheme for broader social MR --- synthesizing realistic avatars that preserve a user's demographic identity while being distinct enough from the individual user to protect facial biometric information.  We introduce a methodology for privatizing appearance by isolating identity within the feature space of identity-encoding generative models.  We develop two algorithms that then obfuscate identity: \epsmethod{} provides differential privacy guarantees and \thetamethod{} provides fine-grained control for the level of identity offset. These methods are shown to successfully generate de-identified virtual avatars across multiple generative architectures in 2D and 3D.  With these techniques, it is possible to protect user privacy while largely preserving attributes related to sense of self.  Employing these techniques in public settings could enable the use of photorealistic avatars broadly in MR, maintaining high realism and immersion without privacy risk.
\end{abstract}

% \begin{IEEEkeywords}
% Virtual / Augmented Reality, user privacy, neural rendering, avatar generation
% \end{IEEEkeywords}

\section{Introduction}

\begin{figure*}[t]
    \centering
    \includegraphics[width=\linewidth]{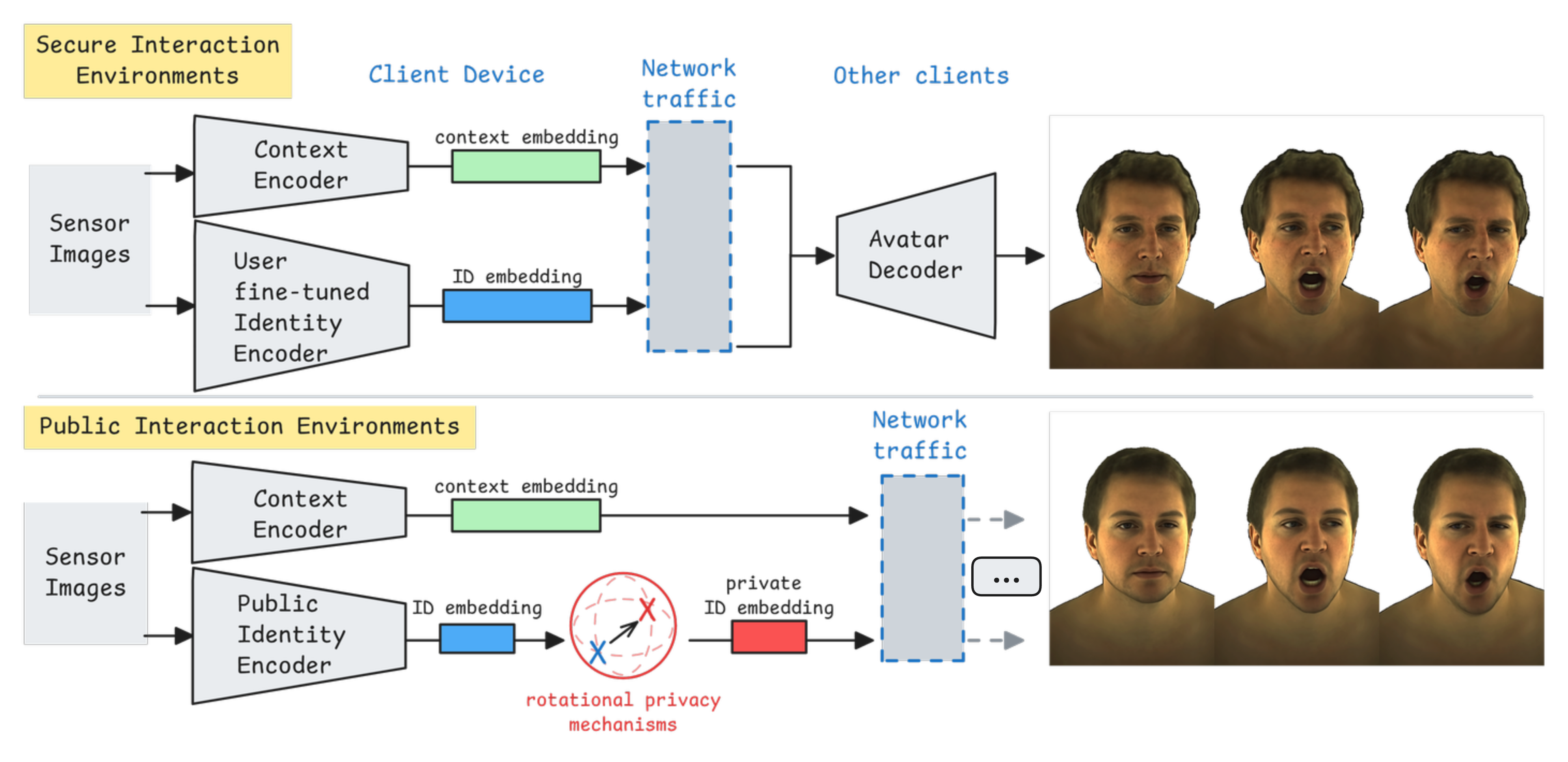}
    \caption{We incorporate a privacy-preserving operation to the photorealistic avatar rendering pipeline when interacting in \textit{Public Interaction Environments}.  In this scheme, user identity embeddings are privatized on-device to protect user appearances while maintaining photorealism and key attributes such as apparent age, race, and gender.}
    \label{fig:teaser_figure}
\end{figure*}

\IEEEPARstart{P}{hotorealistic} 3D avatars have advanced rapidly in recent years, both in the quality achieved and the efficiency of the generation process.  Technical advancements related to avatar generation~\cite{bi_deep_2021, zhi_dual-space_2022, saito_relightable_2024, li_uravatar_2024}, animation~\cite{chen_monogaussianavatar_2024, ma_3d_2024, zhao_media2face_2024}, and rendering efficiency~\cite{saito_squeezeme_2024} now make it plausible to bring photorealistic representations of users into Mixed Reality (MR) and social Metaverse environments.  A number of perceptual evaluations have found realism to improve the quality of social communication~\cite{zibrek_is_2019, aseeri_influence_2021, freeman_body_2021}.  The same technologies could also enable autonomous virtual agents to be created, allowing other users to interact with persistent digital counterparts of MR personalities~\cite{salagean_meeting_2023}.

The impacts of photorealistic self avatars would be broad, and this technology has the potential to improve general social communication~\cite{aseeri_influence_2021}, virtual training scenarios~\cite{xie_review_2021}, healthcare and elderly care~\cite{montenegro_survey_2019}, education~\cite{ripka_preservice_2020}, and more.  However, the broad deployment of photorealistic self avatars poses a number of privacy risks for users.  There are inherent risks associated with sharing personal details online~\cite{acquisti_secrets_2020, kagan_zooming_2024}, and photorealistic self avatars could exacerbate these risks by publicly advertising user identities directly through facial appearance~\cite{agarwal_biometrics_2024}.  MR users' personal details, places of work, addresses, etc. could be easily searched online by linking avatar facial identity to other internet sources~\cite{powar_sok_2023, senette_user_2024}.  Some avatar generation pipelines now streamlined to single photographs~\cite{cao_authentic_2022, li_uravatar_2024}, and this ease of generation makes avatar impersonation attempts inevitable~\cite{tariq_deepfake_2023} in virtual environments that adopt photorealistic avatars.  The risks arising from identity leakage and impersonation can be disproportionately high for child users~\cite{maloney_stay_2021} and those in marginalized groups~\cite{sannon_privacy_2022}.  Thus, proper safeguards are necessary to make social MR interactions with photorealistic avatars viable.

It is necessary to identify and assess these potential privacy risks \textit{before} such avatars are deployed broadly.  Current trends in avatar synthesis are to fine-tune the generation model to the specific user, in order to maximize reconstruction accuracy~\cite{cao_authentic_2022}.  This may be ideal over secure networks while interacting with other known, trusted users, but scaling to broader social scenarios introduces the aforementioned privacy risks surrounding visual appearance leakage.  We propose an alternate privacy-preserving rendering methodology better suited to public environments or third-party servers.  Our proposed solution for \textit{public interaction environments} is to first de-identify avatar appearances on-device before rendering.  This methodology leverages key insights from avatar perception --- In the literature, users have preferred avatars which match their own broad attributes (e.g. aesthetics, gender, race, and perceived age)~\cite{freeman_body_2021}, and users have a self-identification bias for similar looking avatars~\cite{salagean_meeting_2023}.  According to these trends, so long as photorealism is achieved and broad attributes are preserved, visual appearance does not need to match perfectly for users to feel immersed and to have ownership over their avatars. 

% For a privacy-aware user experience, users should be able to adjust the similarity between their avatar and appearance to match a given context.  With the current industry standard for social avatars --- user-chosen stylized / cartoony avatars --- control is possible to an extent by adjusting avatar attributes.  Yet, there is no clear solution for photorealistic avatar rendering beyond switching rendering styles across contexts, though this presents an inconsistent MR experience.  Alternatively, if photorealistic self-avatars become ubiquitous, users would be exposed to significant privacy risk as their true attributes or identity could be broadcast widely without their knowledge or consent.  

To address the privacy concerns of photorealistic self-avatars, we explore methods to generate broadly similar to their users, but adjust avatar identity to enforce some dissimilarity between user and avatar, protecting the privacy of the user's appearance.  This change can help to incorporate photorealistic avatars much more broadly in MR and social Metaverses.  In more intimate scenarios (such as a private lobby of friends enjoying co-presence) and trustworthy applications, avatars rendered with exact likenesses could still be deployed.

This work explores the feasibility of an alternate privacy-preserving rendering schema for photorealistic MR avatars.  To achieve this, we introduce two modular privacy algorithms which can be easily applied to avatar generation models.  Our proposed algorithms leverage a key insight --- once identity is disentangled by the generative model, it can be projected to a high-dimensional directional vector, and identity similarity can be measured with angular metrics.  The first privacy algorithm, \epsmethod{}, operates under a differential privacy framework, providing robust privacy guarantees for user appearances, while allowing tuning of $\varepsilon$ to directly control the privacy-utility trade-off for various appearance attributes.  The second, \thetamethod{}, rotates identity feature vectors a specified amount, enabling user-specified privacy levels for their avatars.  A composition between \epsmethod{} and \thetamethod{} can yield a differentially private, highly tunable privacy solution.

To our knowledge, these are the first methods to achieve privacy-centric rendering of photorealistic 3D avatars.  To better evaluate these algorithms' effectiveness, 2D implementations of our techniques are compared to established 2D facial privacy methods.  Our methodology is implemented on two image-based face synthesis models and one fully 3D avatar generation model. Each implementation effectively privatizes appearances while preserving photorealism, demographic attributes, and utility to a high degree.

Our face anonymization toolkit, which includes the majority of implemented algorithms and the full evaluation pipeline, is available at \url{https://github.com/WahahaYes/FaceAnonEval}.  Extended codebases for the GHOST face synthesis architecture~\cite{groshev_ghostnew_2022} and Relightable Gaussian Codec Avatar architecture~\cite{saito_relightable_2024} which implement our \epsmethod{} and \thetamethod{} algorithms are available at \url{https://github.com/WahahaYes/anonghost} and \url{https://github.com/WahahaYes/ava-256}, respectively.
\section{Related Literature}

\subsection{Realistic Face and Avatar Generation}

The neural rendering of digital humans has become highly detailed and realistic following a number of technical advancements.  Face synthesis and editing has arisen as a popular application space from generative adversarial networks~\cite{gui_review_2023} and diffusion model~\cite{yang_diffusion_2023} architectures.  Face image synthesis is now nearly indistinguishable from real photographs and video~\cite{bozkir_can_2024} and can synthesize novel identities~\cite{karras_analyzing_2020}, condition generated appearances from images of real people~\cite{chen_simswap_2020, groshev_ghostnew_2022}, or adjust real video footage to match driving signals such as audio lip-syncing~\cite{lu_live_2021}. 

With the recent innovations of neural radiance fields (NeRFs)~\cite{yao_neural_2024} and Gaussian splatting~\cite{wu_recent_2024}, rendered 3D virtual avatars have seen large improvements in fidelity and efficiency.~\cite{bi_deep_2021}.  NeRF methods~\cite{zhi_dual-space_2022} allow for the rendering of unseen poses without retraining.  3D Gaussian splatting representations preserve fine details and can be animated in real-time on consumer hardware~\cite{saito_relightable_2024}.  Avatar generation is also becoming more accessible and less costly; 3D representations can be generated from few-shot phone scans~\cite{cao_authentic_2022, li_uravatar_2024}, where in the past, data collection required complex VFX setups.  Additionally, new techniques in avatar animation~\cite{chen_monogaussianavatar_2024, ma_3d_2024, zhao_media2face_2024}, the relighting of avatars to match environmental lighting~\cite{li_uravatar_2024, yang_vrmm_2024}, and rendering pipeline optimizations~\cite{saito_squeezeme_2024} further enhance realism and the feasibility for incorporating photorealistic 3D avatars to MR environments.

\subsubsection{Avatar Representation in Immersive Environments}

Social interactions are a driving force for MR, with known benefits over non-immersive remote telepresence in communication~\cite{aseeri_influence_2021}, education~\cite{ripka_preservice_2020}, healthcare~\cite{shao_acceptance_2020}, and in building social connections or social support~\cite{deighan_social_2023, van_brakel_feelings_2023}.  In social interactions, users wish to be represented consistently and accurately along axes such as aesthetics, gender, race, and age/maturity~\cite{freeman_body_2021}.  Users have also preferred realistic representations when presented with an array of alternatives~\cite{aseeri_influence_2021}, and realism is shown to facilitate stronger emotional connections between users~\cite{zibrek_is_2019}.  The literature indicates that photorealistic avatars which preserve a user's sense of self could be an ideal representation for immersive virtual environments.

\subsection{Appearance Editing}

Many neural rendering methods allow for attribute manipulations.  Various image-based methods isolate and adjust facial features before the generation process~\cite{liu_gan-based_2023}, such as de-aging individuals in photos, imposing new expressions, or adding accessories such as glasses and new hairstyles~\cite{ding_injectiongan_2020}.  Methods which generate target identities~\cite{chen_simswap_2020, groshev_ghostnew_2022} separate identity features from non-identity contextual information (expressions, head pose, gaze directions, etc.).  Recent 3D avatar generation methods have explored the same concepts, allowing attribute editing and stylization through prompting~\cite{liu_headartist_2024} or the automatic generation of 3D caricatures~\cite{jung_deep_2022}.  Some recent 3D avatar generators can synthesize multiple distinct avatars without retraining by using separated identity and non-identity encoders~\cite{saito_relightable_2024, li_uravatar_2024, bai_universal_2024}.

\subsection{Privacy Risks from Appearance Sharing}

There are a number of risks involved when sharing self-information online.  A prominent risk is identity linkage across online sources, which could expose users to phishing attacks~\cite{xu_personalized_2023}, profiling~\cite{marforio_analysis_2012}, and more.  Recorded appearances are especially prone to leaking attributes of interest to adversaries~\cite{jain_online_2021}, and face appearance is a strong positive identifier which enables identity linkage even without the presence of other unique identifiers~\cite{senette_user_2024}.  Privacy risks are disproportionately high for those in marginalized groups~\cite{sannon_privacy_2022}, where exposition of sensitive details could lead to discrimination or harm.

The data collection required to generate realistic user avatars also increases the risks of deepfakes being generated~\cite{zhang_deepfake_2022}.  Adversarial deepfake videos have caused significant harm through scams, creation of non-consensual intimate imagery, and as a tool in political slander~\cite{dobber_microtargeted_2021, busacca_deepfake_2023}.  Deepfakes are expected to persist as a threat in MR through avatar impersonation, and there may be higher risks towards minors and susceptible populations~\cite{tariq_deepfake_2023}.  Because it is possible to generate avatars from cellphone images~\cite{cao_authentic_2022, li_uravatar_2024}, the barrier to entry for avatar impersonation is lowering, as adversaries could bypass the need to acquire encoded avatar features.

\subsection{Privacy-preserving Methods}

Many face synthesis methods have been utilized aiming to share image / video information while protecting user identity.  Traditional approaches, such as blurring, pixelation, and degradation, can no longer protect against advanced recognition systems~\cite{neustaedter_blur_2006, oh_faceless_2016, wilber_can_2016} and are known to degrade utility severely~\cite{li_blur_2017}.  Deep learning approaches for face synthesis can replace or alter an individual's face, yet many of these have been found to leak sensitive information~\cite{mcpherson_defeating_2016, deng_arcface_2019-1} or falter under more rigorous evaluations~\cite{kuang_effective_2021}.  Because such naive privacy operations often become obsolete, privacy research has shifted to prefer principled approaches that offer provable privacy guarantees.  Formulations such as differential privacy (DP) are preferable in developing a robust privacy solution~\cite{dwork_differential_2006, kasiviswanathan_what_2011}.

Many generative methods have been proposed to protect faces in images, including face swapping~\cite{zhu_deepfakes_2020, wilson_practical_2022}, image-based identity feature manipulations~\cite{kuang_effective_2021, zhai_a3gan_2022, xue_face_2023}, and video-consistent manipulations~\cite{maximov_ciagan_2020, cao_achieving_2024}.  These methods protect facial identity against immediate recognition, but lack privacy guarantees, so could succumb to later resourceful adversaries.  The IdentityDP method proposes a local differential privacy (LDP) operation, adding 1-D Laplace noise samples to embedded identity features when generating face images~\cite{wen_identitydp_2022}.  However, IdentityDP's assumptions are violated when adapted to 3D avatar generation, resulting in unintelligible output (discussed in \autoref{sec:comp_identitydp}).

Prior to this work, privacy-preserving approaches for 3D facial avatar generation had not been explored.  If photorealistic self-avatars were used ubiquitously in social Metaverses, MR systems would rely entirely on robust authentication~\cite{lohr_demonstrating_2023, yang_secure_2023} to prevent avatar impersonation attempts.  On top of constraining photorealistic MR to secured first-party applications, this induces a reliance on application infrastructures, and does not grant individual users control over their own privacy~\cite{acquisti_secrets_2020}.  Lingering challenges would remain unsolved, such as how to prevent users' appearances from simply being screen captured without consent.
\section{Motivation}

There are risks associated with publicly sharing personal details online~\cite{acquisti_secrets_2020, zhang_privacy_2010, xu_personalized_2023, marforio_analysis_2012}.  Photorealistic virtual self-avatars potentially exacerbate these risks by attaching a permanent, persistent identifier (facial appearance) to users' online actions~\cite{agarwal_biometrics_2024, kagan_zooming_2024, jain_online_2021}.  Online users' personal details, place of work, addresses, etc. could be learned by linking facial identity to other internet sources~\cite{powar_sok_2023, senette_user_2024}.  Thus, proper safeguards are necessary to protect users' online privacy, and to make social MR interactions with photorealistic avatars viable.

We envision a new design scheme for photorealistic social MR consisting of a \textit{public interaction environment} and \textit{secure interaction environment}, depending on the application context and security of the system.  In secure interaction environments, user identifying features are protected by the application.  In public interaction environments, user identifying features are protected by being altered on-device, such that networked peers are unable to access the user's true appearance information at any stage of the interaction.  The difference in processing of avatar information between secure / public interaction environments is illustrated in \autoref{fig:teaser_figure}.

\subsection{Secure Interaction Environments}

The context which we coin as a secure interaction environment is aligned with the current trajectory for photorealistic MR (see \autoref{fig:teaser_figure} (top)).  In this environment, users' avatars would appear as faithful of a reconstruction as possible, likely through model fine-tuning~\cite{cao_authentic_2022}.  This maximizes social presence but requires multiple layers of security to prevent compromising avatar likenesses.  User likenesses must be enrolled on a authentication server~\cite{lohr_demonstrating_2023, yang_secure_2023} and usage must be restricted to secure first-party applications to preserve the integrity of identity features.  This environment should be constrained to trusted social connections, such as family members or authenticated coworkers, to prevent unwanted identity lookup from malicious strangers~\cite{senette_user_2024}. 

\subsection{Public Interaction Environments}

While the secure interaction environment is well-suited for intimate interactions, such as virtual calling, it does not scale well to public scenarios, such as social gaming or large group hangouts.  As a concrete example, imagine a support group, such as an \textit{Alcoholics Anonymous} meeting, in a private room on a virtual platform and using the platform's photorealistic self avatars.  Despite the anonymity posed by the group, a malicious participant could lookup users by their appearances, associate their real-world identity with the group's sensitive attribute(s), then exploit this knowledge.  Additionally, if third party applications were permitted to render avatars, raw identity features --- which can be considered a biometric~\cite{agarwal_biometrics_2024} --- could be harvested if application permissions were misused.

To better facilitate photorealistic avatars across MR, we propose rendering de-identified avatars when in public interaction environments.  This can mitigate the risk of identity leakage when interfacing with potentially untrustworthy systems or peers.  This can be achieved by altering avatars to have distinct appearances while maintaining attribute consistency and photorealism (see \autoref{fig:teaser_figure} (bottom)).  Thus, privacy can be granted while preserving heightened immersion and sense of self~\cite{zibrek_is_2019, aseeri_influence_2021, freeman_body_2021}.

\subsection{Contributions}

This work articulates the need for a privacy-preserving rendering scheme when users are represented as photorealistic avatars and proposes technical solutions to achieve this goal.  Two privacy-preserving algorithms are developed --- \epsmethod{} is formulated to provide LDP guarantees, and \thetamethod{} allows specified dissimilarity of identities at the individual sampling level.  The resulting avatars are similar, with parameterized similarities to original users, but are distinctly randomized, preventing identification of the MR user.  An application may incorporate a composition between \epsmethod{} and \thetamethod{} to yield a differentially private, highly tunable privacy solution for the rendering of user avatars.

These algorithms can be adapted to existing generative models and could generalize to other application areas which synthesize user representations.  While a number of recent approaches operate on encoded identity features~\cite{wen_identitydp_2022, xue_face_2023, maximov_ciagan_2020}, generative models tend to embed into arbitrary spaces, so it is difficult to consistently produce dissimilar outputs.  Our approach incorporates an operation to remap features into a more standardized domain, mitigating the challenge of randomizing arbitrary embeddings.  In this work, the proposed methodology is applied across 3D avatar and 2D face synthesis models, showcasing a strong ability to generalize across generative AI architectures with acceptable privacy-utility trade-offs.  

The rest of the paper is structured as follows.  \autoref{sec:threat_model} details the threat model for envisioned MR adversaries which wish to exploit user appearances.  \autoref{sec:preliminaries} covers mathematical preliminaries needed to interpret the algorithmic details of our proposed privacy mechanisms.  \autoref{sec:methodology} details the implemented avatar generation models and the components of our evaluation pipeline.  \autoref{sec:results} covers the results of our privacy-utility evaluations.  \autoref{sec:discussion} presents a discussion of our findings, limitations, and future research avenues.

\section{Threat Model}
\label{sec:threat_model}

\begin{figure*}[h!]
    \centering
    \includegraphics[width=\linewidth]{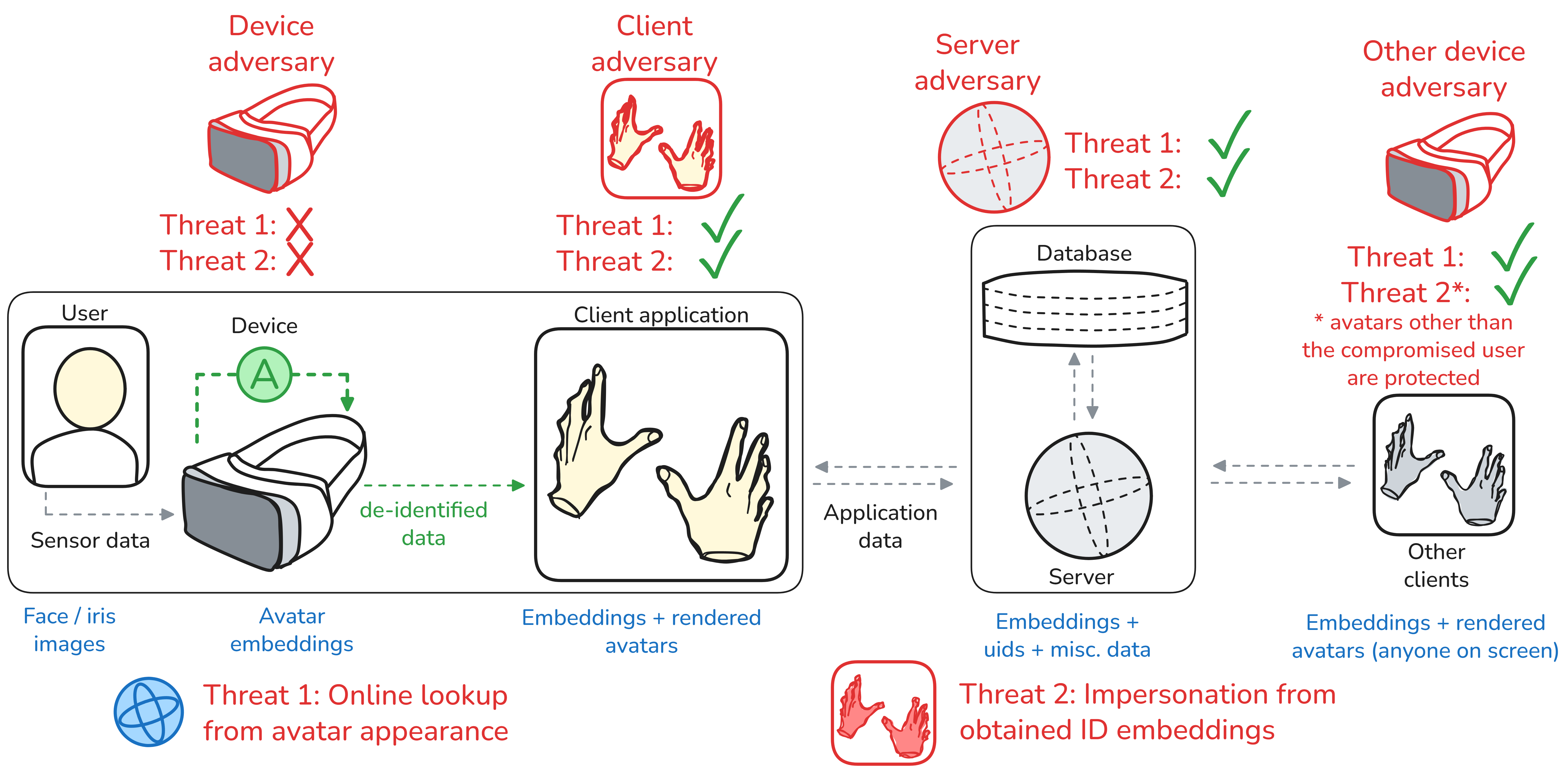}
    \caption{Illustration of the network communications involved in social MR, the avatar-related information being shared, and where various adversarial threats exist.  Our device-level privacy mechanisms protect against client, server, and other-device adversaries by obfuscating avatar appearance data on-device before sending to any external system for rendering or storage.  Checkmarks indicate that the proposed privacy-preserving operations ($\mathcal{A}$) protect users from a given threat and adversary.}
    \label{fig:threat_model}
\end{figure*}

When discussing privacy threats, a standard assumption in privacy literature is that adversaries are knowledgeable of the avatar generation process and which privacy mechanisms are being used, but that they do not know values of random variables.  Our threat model operates under this assumption.  Adversaries can exploit users by recording their rendered appearance (using these appearances for unwanted lookup or profiling) or by storing their encoded identity features to falsely generate users for impersonation.  This section discusses adversaries which could feasibly be encountered.  Threats include malicious applications running on-device, untrustworthy or compromised servers / data centers, and when other client devices on the network are compromised.  \autoref{fig:threat_model} shows where on a MR network each adversarial threat may exist.

\subsection{Untrustworthy Applications}

Untrustworthy applications, which could appear benign to users, may aim to store user information and link these records to other internet sources.  These applications could extract sensitive user details via adversarial game design~\cite{nair_exploring_2023} or other means and attach said details to the user likenesses extracted from their avatars.  Applications could additionally misuse the encoded identity features meant for avatar rendering, such as storing these features to use for later impersonation.

\subsection{Malicious Users}

Malicious users may encounter the target user in MR environments.  By recording a user's appearance (possibly with a simple screenshot), they are able look up the target user's real information from online sources~\cite{senette_user_2024}.  Later, this information could be exploited for blackmail or cyber abuse.  

A user may unknowingly aid an attack if their device or application client were compromised.  Without privacy protections, the compromised device which receives encoded avatar features would be able to harvest said features (possibly uploading to another network) for later impersonation or identity lookup.

\subsection{Device Security Considerations}

If a user's MR device were compromised, the raw eye and face sensor captures would enable identification and avatar reconstruction, among additional risks.  Thus, device security is out of scope for the privacy protections being proposed, and hardware protections are necessary through other means~\cite{giaretta_security_2024}.  Device-level user authentication would remain necessary for authenticating into secure interaction environments~\cite{lohr_demonstrating_2023}.
\section{Preliminaries}
\label{sec:preliminaries}

The proposed privacy mechanisms leverage the fact that identity-encoding generative models disentangle \textit{identity} information from \textit{contextual} information during the generation process, enabling the manipulation of \textit{identity} with minimal impact to other features.  Identity disentanglement is not perfect, however, and improved disentanglement is an active area of research~\cite{shiohara_blendface_2023, choi_dddm-vc_2024}.

During encoding, the majority of generative models embed identity features to high-dimensional spaces.  The proposed privacy mechanisms assume a disentangled identity and a high-dimensional embedding space.  \epsmethod{} is based on differential privacy formulations, and \thetamethod{} performs a targeted rotation.  In this section, definitions are given for the necessary concepts of differential privacy operations and of high-dimensional geometric spaces.  

% \subsection{Avatar Embedding Assumptions}

% \todo{(8) Here would be a good place to talk about embeddings and how the facial/avatar models work and what they assume about the embedding space. I would even suggest talking about this before talking about differential privacy.\\
% \\
% Short section of how identity embeddings are produced and any assumptions (arcface gives you an angular embedding, others do not).  Typically ML uses soft constraints rather than hard constraints, so even wanted behaviors are somewhat unreliable}

\subsection{Differential Privacy}

Differential privacy~\cite{dwork_differential_2006} is a mathematical formulation for noisy data release which bounds the impact that a single data record can have when either included or excluded from any dataset.  DP is a gold standard for privacy-preserving operations, being robust against an adversary's auxiliary knowledge and having multiple favorable properties.  These properties include sequential composition~\cite{dwork_our_2006, dwork_calibrating_2006}, parallel composition~\cite{mcsherry_privacy_2009}, and a post-processing guarantee~\cite{bun_concentrated_2016} where future operations on anonymized data preserve the DP guarantees.

DP is designed for protecting the inclusion and exclusion of data records in full datasets. Local differential privacy (LDP)~\cite{kasiviswanathan_what_2011} is a variant which protects privacy at the granularity of an individual's data (\autoref{def:ldp}).  The LDP formulation is ideal for appearance anonymization, providing privacy guarantees for each individual user's rendered appearance, without relying on a trusted data aggregator.

\begin{definition}[$\varepsilon$ Local Differential Privacy]
\label{def:ldp}
For any two records $x_1, x_2$ existing within a shared domain $\mathcal{X}$, a randomized algorithm $\mathcal{A}$ satisfies $\varepsilon$ local differential privacy if for all valid subsets of possible outcomes $\mathcal{S} \in Range(\mathcal{A})$:
\[ \Pr [\mathcal{A}(x_1) \in \mathcal{S}] \leq e^{\varepsilon} \Pr [\mathcal{A}(x_2) \in \mathcal{S}] \]
\end{definition}

A popular method to satisfy DP / LDP for numerical data is the Laplace mechanism~\cite{dwork_calibrating_2006}, which samples random noise from the Laplace distribution \textit{Lap}$(x \ ; \ \mu, b) = \frac{1}{2b} e^{- \frac{\vert x - \mu \vert}{b}}$ according to the L$_1$ sensitivity of the dataset or domain (\autoref{def:l1_sensitivity}).

\begin{definition}[L$_1$ Sensitivity]
\label{def:l1_sensitivity}
For domain $\mathcal{X}$ and query function $f: \mathcal{X} \rightarrow \mathbb{R}$, the L$_1$ sensitivity of $f$ is:
\[ \triangle f = \max_{x_1, x_2} \vert f(x_1) - f(x_2) \vert_1 \qquad \forall \{x_1, x_2\} \in \mathcal{X} \]  
\end{definition}

\subsection{Metric Privacy}

Metric privacy is a variant of LDP with relaxed constraints by weighing the relative distance between compared samples.  The probability ratio between two privatized records $x, x'$ is dependent on the measured distance $d_{\mathcal{X}}(x, x')$, subject to a given metric $d_{\mathcal{X}}$~\cite{alvim_invited_2018}.

\begin{definition}[Metric $d_{\mathcal{X}}$]
A metric on an arbitrary set $\mathcal{V}$ is a function $d_{\mathcal{X}}: \mathcal{V}^2 \rightarrow [0, \infty]$ such that $d_{\mathcal{X}}(x, y) = 0$ iff $x \equiv y$, $d_{\mathcal{X}}(x, y) = d_{\mathcal{X}}(y, x)$, and $d_{\mathcal{X}}(x, z) \leq d_{\mathcal{X}}(x, y) + d_{\mathcal{X}}(y, z)$ for all $x, y, z \in \mathcal{V}$.
\end{definition}

\begin{definition}[Metric Privacy / $d_{\mathcal{X}}$ Privacy]
A randomized algorithm $A: \mathcal{V} \rightarrow \mathbb{R}$ is $d_{\mathcal{X}}$ private if for every pair of inputs $x_1, x_2 \in \mathcal{V}$ and for all valid subsets of possible outcomes $\mathcal{S} \in \mathcal{A}$: 
\[ \Pr [\mathcal{A}(x_1) \in \mathcal{S}] \leq e^{d_{\mathcal{X}}(x_1, x_2)} \Pr [\mathcal{A}(x_2) \in \mathcal{S}] \]
\end{definition}

\subsection{Hypersphere Representation}

In our methods, we map identity embeddings $x_{id}$ produced from generative models to directional vectors equivalent to points on the surface of high-dimensional hyperspheres $\mathbb{S}^{n-1}$ (see \autoref{def:unit_sphere}).  The angular distance $\theta$ between different samples on $\mathbb{S}^{n-1}$ is used as metric $d_{\angle}$.  

\begin{definition}[Unit Hypersphere $\mathbb{S}^{n-1}$]
\label{def:unit_sphere}
For $n \in \mathbb{N}$, the unit hypersphere $\mathbb{S}^{n-1} = \{x \in \mathbb{R}^n : \vert x \vert_2 = 1\}$ is the set of unit vectors in $\mathbb{R}^n$ and has surface area:
\[ S_{n - 1} = \frac{2\pi^{\frac{n}{2}}}{\Gamma(\frac{n}{2})} \] 
Where $\Gamma$ is the Gamma function.  
\end{definition}

The proposed \epsmethod{} resamples new points from $\mathbb{S}^{n-1}$, using the von Mises-Fisher (VMF) distribution of $\mathbb{R}^{n}$, where $n = dim(x_{id})$ ~\cite{wood_simulation_1994}.  The VMF distribution aligns around a mean direction $\mu$ and concentration parameter $k$ to draw random samples (\autoref{def:vmf}).  Weggenmann and Kerschbaum proved the VMF distribution is a valid mechanism to satisfy $\varepsilon$ LDP and $\varepsilon d_{\angle}$ privacy~\cite{weggenmann_differential_2021}.

\begin{definition}[von Mises-Fisher Distribution]
\label{def:vmf}
For mean direction $\mu \in \mathbb{S}^{n-1}$ and concentration parameter $k \geq 0$, the probability density function of the VMF distribution is:
\[ VMF(\mu, k) \ (x) = \frac{k^{n/2 - 1}}{2 \pi^{n/2} I_{n/2 - 1}(k) } e^{k \mu \cdot x} \]
where $I_a$ is the modified Bessel function of the first kind.
\end{definition}

The proposed \epsmethod{} resamples encoded $x_{id}$ according to the VMF mechanism, satisfying $\varepsilon d_{\angle}$ privacy and $\varepsilon$ LDP, and \thetamethod{} rotates $x_{id}$ along a random direction with a target magnitude $\theta$, ensuring that $d_{\angle} (x, x') \equiv \theta$.  While not rooted in probability theory, \thetamethod{} has some adversarial uncertainty induced from the random direction component of the rotation, while offering consistent levels of change for individual resamples.

\subsection{Privacy Guarantees of the von Mises-Fisher Distribution}
\label{sec:privacy_of_vmf}

Here we briefly overview Weggenmann and Kershbaum's proof of the $\varepsilon d_{\angle}$ privacy of the VMF distribution, in consistent notation with ours.  Weggenmann and Kerschbaum proved that the VMF mechanism can satisfy differential privacy by relating the sensitivities of $\varepsilon d_{\mathcal{X}}$ and $\varepsilon$ LDP methods (\autoref{thrm:implies}).  A synopsis of their proof is as follows.  They prove the $d_{\mathcal{X}}$ privacy of L$_2$ distance $d_2 (x_1, x_2)$ (\autoref{thrm:d2}).  From there, it is simple to substitute $d_{\angle}$ which is $\leq d_2$ at all points (\autoref{cor:d_angle_VMF})~\cite{weggenmann_differential_2021}.

\begin{theorem}[$\varepsilon d_{\mathcal{X}}$ Privacy Implies $\varepsilon$ Differential Privacy]
\label{thrm:implies}
Let $f: \mathcal{V} \rightarrow \mathbb{S}^{n-1}$ be a query function, and let $\mathcal{A}_{\varepsilon}$ be a $\varepsilon d_{\mathcal{X}}$ private mechanism with metric $d_{\mathcal{X}}$.  The sensitivity of $\mathcal{A}_{\varepsilon}$ is:
\[ \triangle = \triangle_{d_{\mathcal{X}}}f = \max_{x_1, x_2 \in \mathcal{V}} d_{\mathcal{X}}(f(x_1), f(x_2)) \] 
\[ \textrm{and the composition } \mathcal{A}_{\frac{\varepsilon}{\triangle}} \circ f \textrm{ is $\varepsilon$-differentially private.} \]
\end{theorem}

\begin{theorem}[$\varepsilon d_2$ Privacy of the VMF Mechanism]
\label{thrm:d2}
Let $\varepsilon > 0$ be a privacy parameter.  The VMF mechanism on $\mathbb{S}^{n-1}$ induced by $x \mapsto$ VMF$(x, \varepsilon)$ for $x \in \mathbb{S}^{n-1}$ fulfills $\varepsilon d_2$ privacy.
\end{theorem}

\begin{corollary}[$\varepsilon d_{\angle}$ Privacy of the VMF Mechanism]
\label{cor:d_angle_VMF}
For any $x_1, x_2 \in \mathbb{S}^{n-1}$, $d_{\angle}(x_1, x_2) \leq d_2 (x_1, x_2)$.  By \autoref{thrm:implies}, the VMF mechanism also implies $\varepsilon d_{\angle}$ privacy.
\end{corollary}
\section{Methodology}
\label{sec:methodology}

\begin{figure*}[t]
    \centering
    \includegraphics[width=0.75\linewidth]{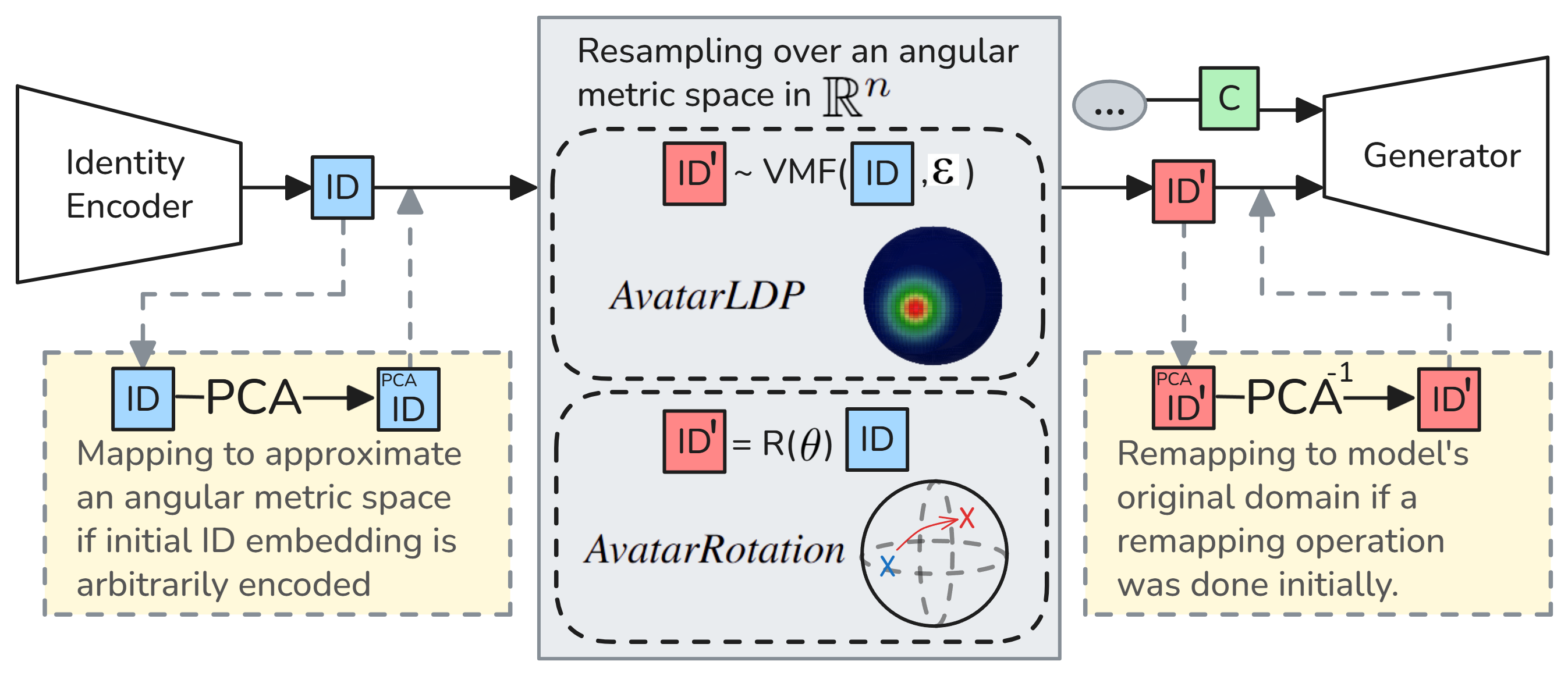}
    \caption{Overview of the processing for encoded identity features.  \epsmethod{} samples a new identity via a von Mises-Fischer distribution conditioned on the identity encoding (ID) and $\varepsilon$.  \thetamethod{} rotates the input identity with a rotation matrix $R$ with random direction and magnitude $\theta$.  For models that encode identity into arbitrary spaces, ID is mapped to a condensed representation using \textbf{PCA} to better model an angular metric space, then remapped to the model's original domain after privatization.  Finally the manipulated ID and contextual encoding (C) together generate the resulting avatar.}
    \label{fig:id_manipulation_diagram}
\end{figure*}

\subsection{Privacy Mechanisms}

The developed privacy mechanisms operate on the embeddings computed by an identity encoder ($x_{id}$) midway through inference of an identity-encoding generative model.  Assuming that identity has been effectively disentangled from other attributes, the features that encode sensitive identity information are perturbed, privatizing the full output of the generative model.  Attributes contained in non-identity features are minimally altered (for example, head pose, gaze direction, expressions, lighting, etc. are preserved).  We present two methods operating on encoded identity features; the first satisfies LDP guarantees (\epsmethod{}), and the second offers a targeted rotation to achieve a specified amount of change (\thetamethod{}).  The identity processing pipeline is visualized in \autoref{fig:id_manipulation_diagram}.

\epsmethod{} provides LDP guarantees using a rotational metric privacy formulation~\cite{weggenmann_differential_2021}.  Given $x_{id}$ encoded into $n$ dimensions, a random sample is drawn from the von Mises-Fisher distribution, centered on $x_{id}$.  The privatized sample $\tilde{x}_{id}$ is drawn from the following probability density function:

\begin{equation}
VMF(x_{id}, \varepsilon) \ (\tilde{x}_{id}) = \frac{\varepsilon^{n/2-1}}{2 \pi^{n/2} I_{n/2-1}(\varepsilon) } e^{\varepsilon x_{id} \cdot \tilde{x}_{id}} 
\end{equation}

\begin{figure}[h]
    \centering
    \includegraphics[width=\linewidth]{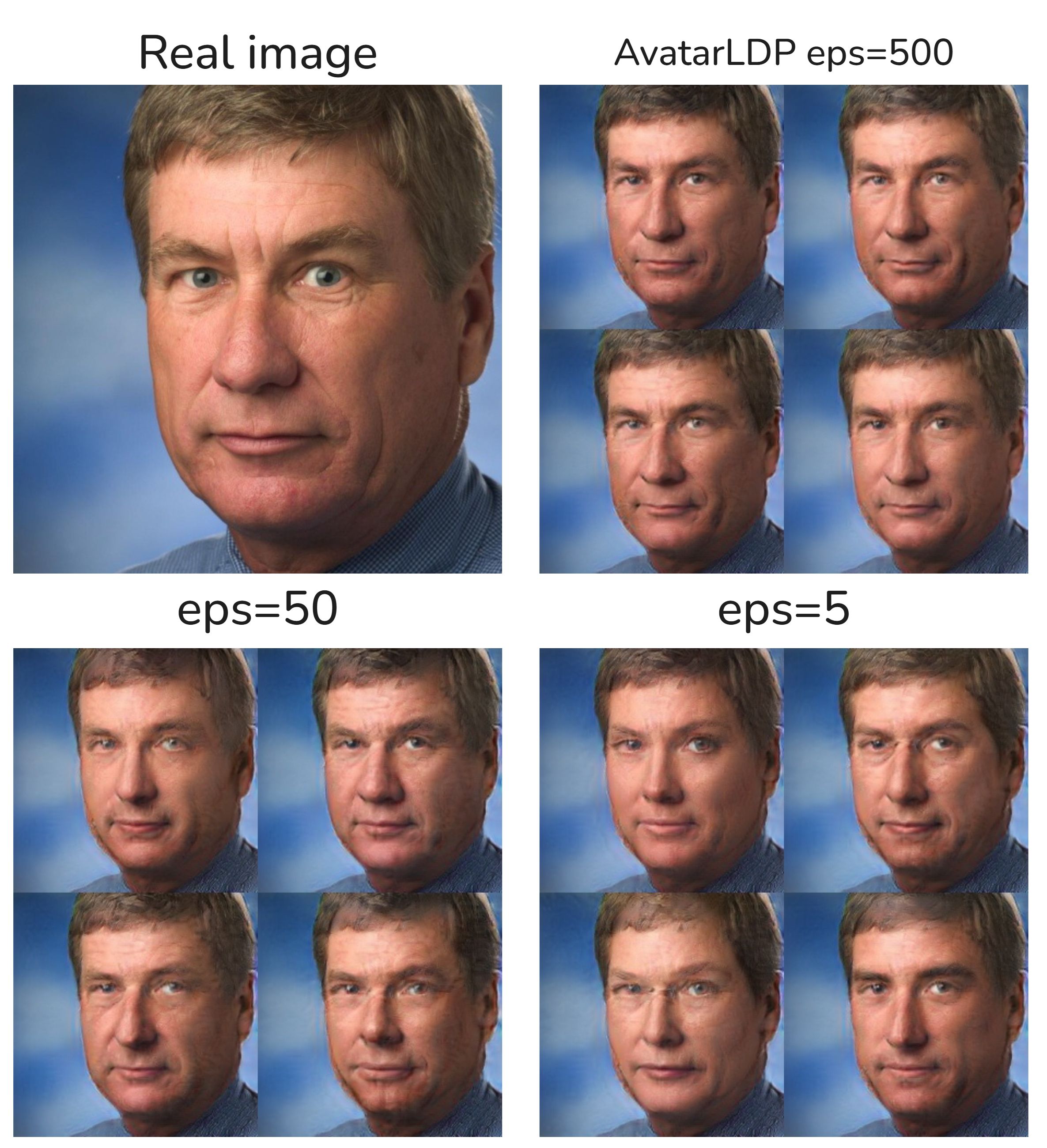}
    \caption{Visualization of the variety of appearances generated by \epsmethod{}'s randomized sampling.  Identities become more dissimilar, on average, as $\varepsilon$ decreases.}
    \label{fig:ldp_variety_results}
\end{figure}

The sampled $\tilde{x}_{id}$ satisfies $\varepsilon$ LDP as proven by Weggenmann and Kerschbaum~\cite{weggenmann_differential_2021}.  As $\varepsilon \rightarrow 0$, the sampling method approaches uniform random sampling over $\mathbb{S}^{n-1}$.  As $\varepsilon$ increases, the probability density tightens around $x_{id}$.  While a handful of DP mechanisms have been developed targeting the embedding space of generative models~\cite{wen_identitydp_2022, weggenmann_dp-vae_2022}, \epsmethod{} uniquely leverages the high-dimensional properties of features embedded into metric spaces.  The randomized variety of \epsmethod{}'s sampled identity vectors can be seen in \autoref{fig:ldp_variety_results}.

In commercial applications, there is a preference among users for user-driven privacy controls~\cite{acquisti_secrets_2020}, so the individual sample randomness of DP methods may be unintuitive and lead to consumer confusion.  An adversary can not verify which records are altered, or how severely, but end-users may perceive minimally perturbed results as a breach in their own privacy.  

To better serve use cases the require \textbf{guaranteed dissimilarity}, \thetamethod{} is proposed as an alternative.  Identity record $x_{id}$ rotates along $\mathbb{S}^{n-1}$ in a random direction with specified magnitude $\theta$.  This is done by sampling a random vector $\{z \in \mathbb{R}^{n} : \vert z \vert_2 = 1 \}$ to form a random orthonormal basis between $z$ and $x_{id}$ using singular value decomposition (SVD).  

\begin{equation}
\begin{split}
A = \begin{bmatrix} {x}_{id} & z \end{bmatrix} & = U \Sigma V^{T} \\
\begin{bmatrix} b_1 & b_2 \end{bmatrix} & = \frac{U}{\vert U \vert}
\end{split}
\end{equation}

Using Rodrigues' rotation formula~\cite{dai_eulerrodrigues_2015} expanded into higher dimensions, the rotation matrix $R$ is computed as:

\begin{equation}
R = I^{n} + (b_2 b_1^T - b_1 b_2^T)\sin\theta + (b_1 b_1^T + b_2 b_2^T)(\cos\theta - 1)
\end{equation}

Finally, $x_{id}$ is rotated to yield $x_{id}'$; $x_{id}' = R \ x_{id}$.  When computing $x_{id}'$, the magnitude $\theta$ is fixed and the direction of rotation $R$ is random.  The uncertainty from the random rotation makes the method impossible to reverse without repeated observations, even if adversaries obtain access to model weights and $\theta$ values.  The scaling of $\theta$ produces non-linear appearance changes because random samples are concentrated around $90\degree$.  $\theta > 90\degree$ enforces dissimilarity beyond what is achievable from random sampling.

\subsection{Generative Model Implementations}

\begin{figure*}[t]
    \centering
    \includegraphics[width=\linewidth]{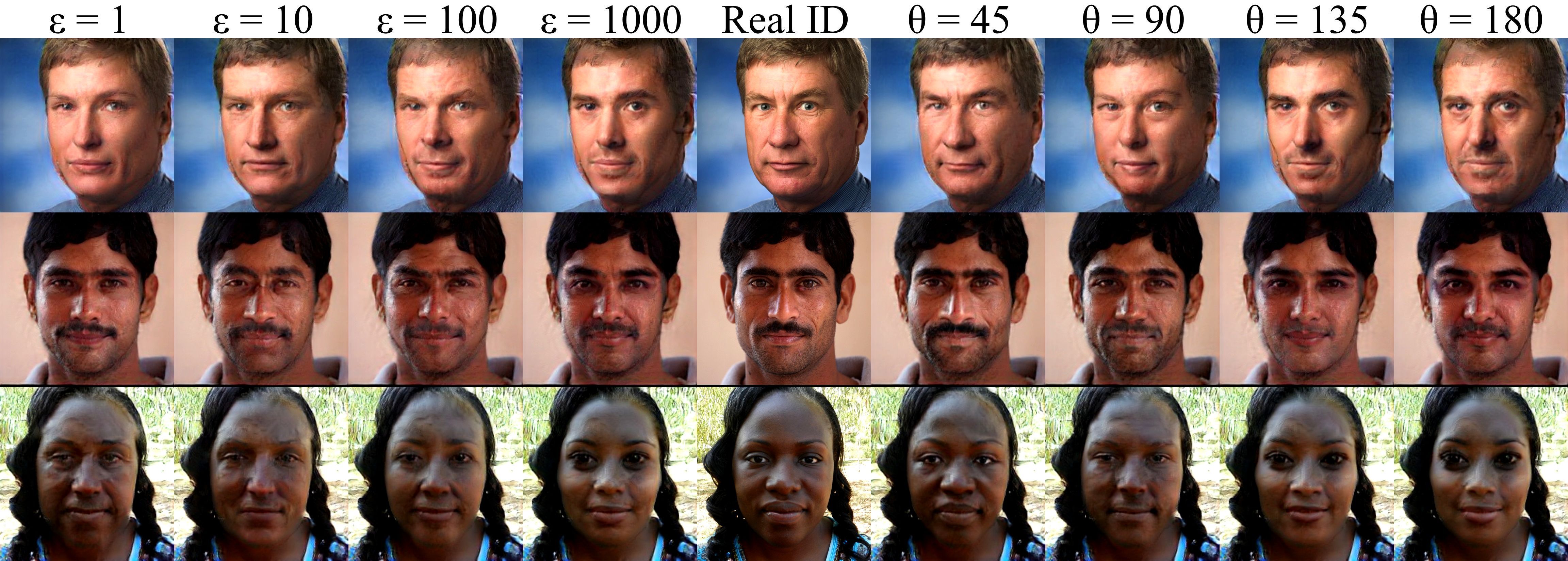}
    \caption{Faces generated via 2D face synthesis using the SimSwap architecture.  Identity embeddings further from the real identity (low $\varepsilon$; high $\theta$) produce more dissimilar results.  Random variables are fixed for visualization purposes.}
    \label{fig:simswap_results}
\end{figure*}

\begin{figure*}[t]
    \centering
    \includegraphics[width=\linewidth]{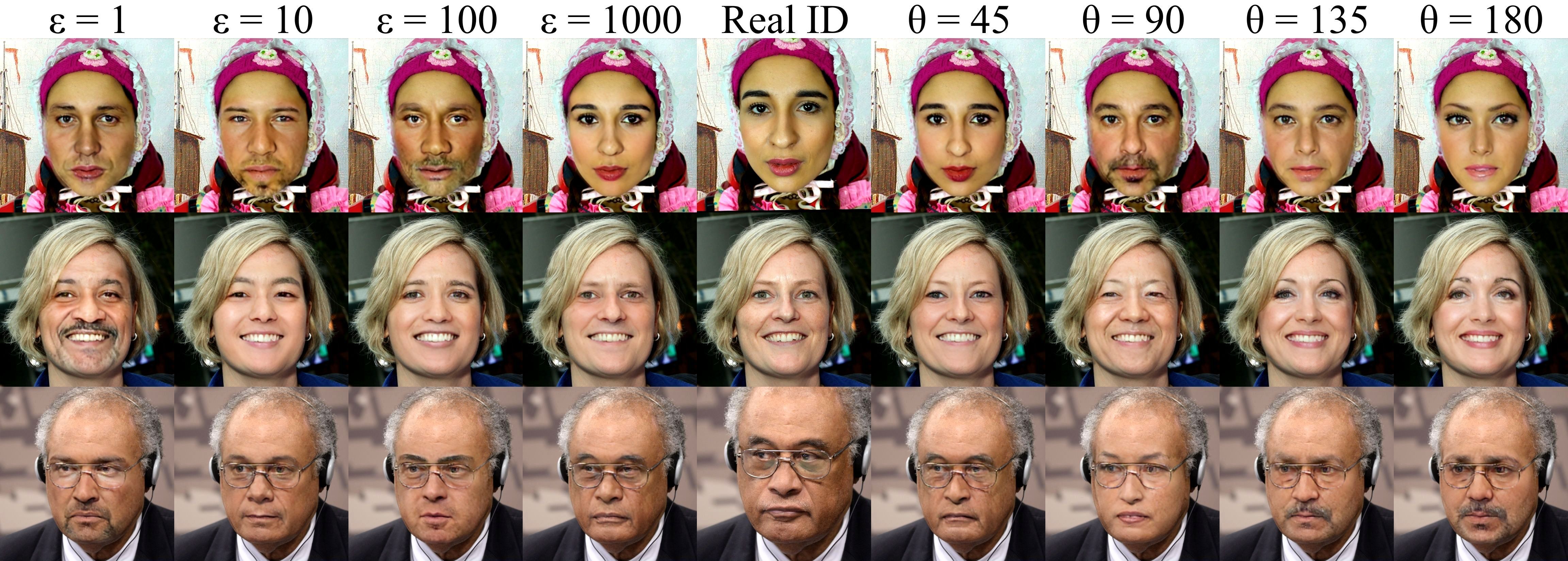}
    \caption{Faces generated via 2D face synthesis using the GHOST architecture, presented in consistent format with \autoref{fig:simswap_results}.  Video results are provided as supplementary material.}
    \label{fig:ghost_results}
\end{figure*}

SimSwap\footnote{SimSwap code: \url{https://github.com/neuralchen/SimSwap}}~\cite{chen_simswap_2020} and GHOST\footnote{GHOST code: \url{https://github.com/ai-forever/ghost}}~\cite{groshev_ghostnew_2022} are two face synthesis models used for evaluating the developed privacy mechanisms.  Both models have the shared purpose of zero-shot synthesis to inject any new identity onto a source image without retraining.  SimSwap uses style transfer~\cite{huang_arbitrary_2017} to replace identity features within the bottleneck of an autoencoder, and GHOST progressively infuses identity into the feature maps a a U-net trained on contextual attributes~\cite{ronneberger_u-net_2015}.  Both methods use pretrained Arcface encoders for identity features~\cite{deng_arcface_2019}.  Privatized faces created with SimSwap are visualized in \autoref{fig:simswap_results}, and GHOST is visualized in \autoref{fig:ghost_results}.

\begin{figure}[h]
    \centering
    \includegraphics[width=\linewidth]{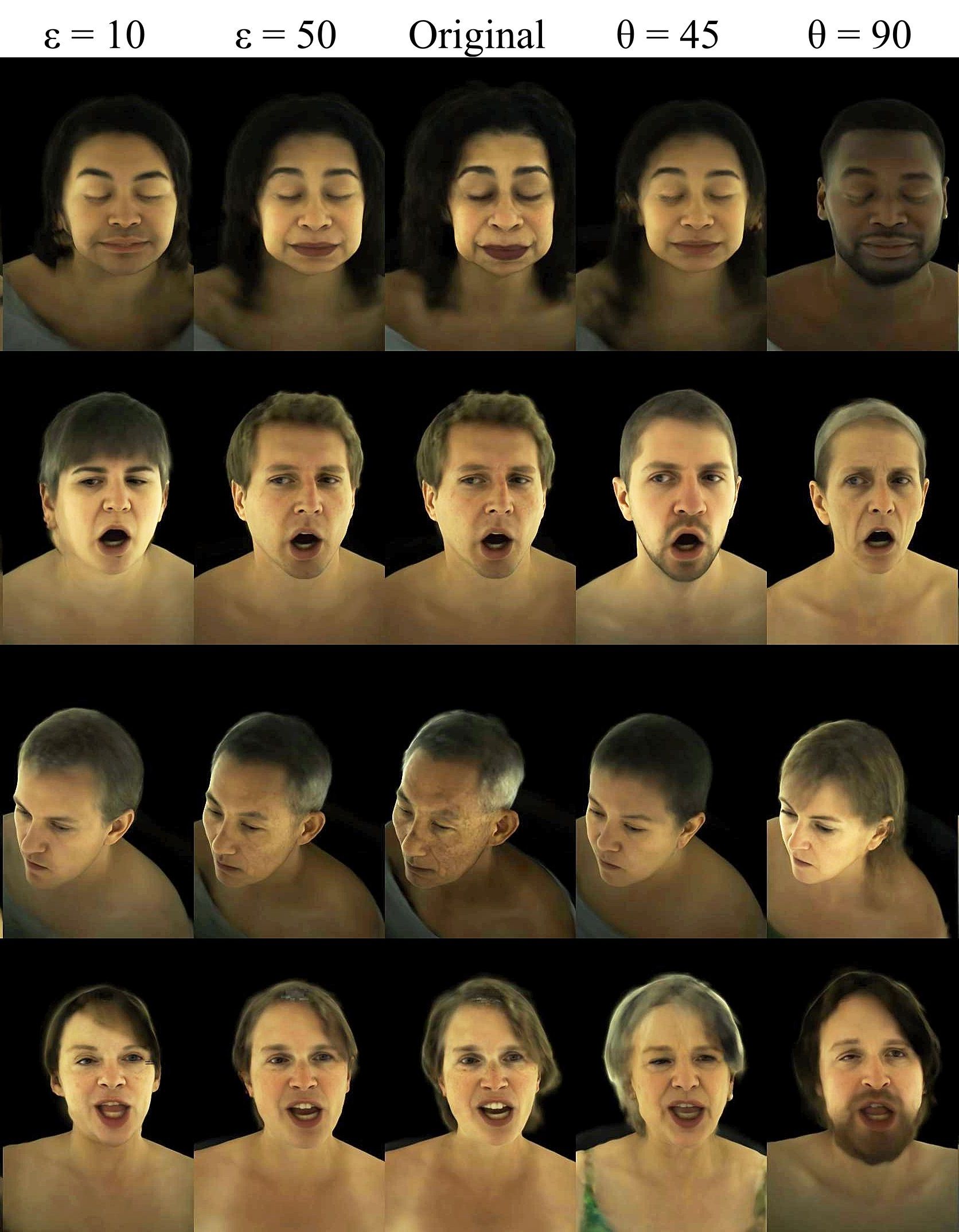}
    \caption{3D avatars generated with the codec avatar architecture.  Identity embeddings further from the real identity (low $\varepsilon$; high $\theta$) produce more dissimilar results.  Video results are provided as supplementary material.}
    \label{fig:avatar_results}
\end{figure}

For 3D avatar generation, the Relightable Gaussian Codec Avatars architecture is used~\cite{saito_relightable_2024}\footnote{Codec avatar codebase: \url{https://github.com/facebookresearch/ava-256}~\cite{martinez_codec_2024}}.  Privatized 3D codec avatars are seen in \autoref{fig:avatar_results}.  The released training dataset has few identities ($N = 256$), and the trained model produces sparse identity embeddings.  As such, identity embeddings must be remapped to a suitable angular metric space to satisfy the requirements of our privacy mechanisms.

\subsection{Remapping Identity Embeddings}

A requirement for an effective privacy mechanism is the ability to map its outputs into a plausible domain.  Given a randomizing privacy algorithm $\mathcal{A}$, if transposing data records within the same domain, then $\forall \ x \in \mathcal{X}, \ \mathcal{A}(x) \in \mathcal{X}$.  This assumption can be satisfied for continuous domains with relative ease. However, in sparsely populated domains, additive noise could produce output values that do not overlap with the input domain.

Autoencoder generative models first compress information into vector representations~\cite{baldi_autoencoders_2012}.  Feature embeddings preserve key information about the input, and the embedding format is dependent on the model architecture and training strategies.  There are known strategies to embed continuous feature representations~\cite{kingma_auto-encoding_2022, gulrajani_improved_2017, deng_arcface_2019}; many recent face synthesis approaches have used Arcface embeddings to represent identity features on a manifold when generating faces~\cite{chen_simswap_2020, groshev_ghostnew_2022}.  But, many other architectures lack the need for these constraints.  In some cases, sparsity is even preferred; Cao et al.'s codec avatar identity encoder deliberately omitted a continuous latent space to prevent a loss of detail from generalization~\cite{cao_authentic_2022}s.  

When angular distance $d_{\angle}$ expresses the relationship between embeddings, embeddings can be projected to points on the hypersphere $\mathbb{S}^{n-1}$.  This hyperspherical representation forms the basis for the proposed privacy mechanisms, and is readily available when identity is encoded using Arcface~\cite{deng_arcface_2019} or a similar approach.  

The 3D codec avatar model implemented in this work encodes identity using U-net architectures~\cite{cao_authentic_2022, ronneberger_u-net_2015} for texture and geometry data, respectively.  Simple resampling results in unintelligible output due to sparsity of embeddings.  The architecture could be tweaked and retrained using various ML techniques to re-encode information as an angular metric space; however, due to the growing size and quantity of large foundation models, expensive retraining iterations may not be feasible.  For this case, we investigate remapping the pretrained model's identity embeddings to what can be interpreted as an angular metric space.

The codec avatar model's texture features, denoted as $x_{id}^{{}^{tex}}$ and occupying $\mathbb{R}^{256}$, are first encoded for all 256 training identities.  Using principal component analysis (PCA) with data standardization, features are remapped to a densely populated set of embeddings in $\mathbb{R}^{16}$.  Here, privacy operations based on $\mathbb{S}^{n-1}$ can be applied effectively, because the relationship between samples is expressed largely through angular metrics.  After resampling in $\mathbb{R}^{16}$, the original format in $\mathbb{R}^{256}$ can be reconstructed, resulting in an anonymized output embedding which exists in the space of embeddings interpretable by the generative model, despite the original embedding set being sparsely populated.   

\begin{equation}
\begin{split}
x_{id}^{{}^{PCA}} & = \ PCA \ ( \ \frac{x_{id}^{{}^{tex}} - mean(x_{id}^{{}^{tex}})}{\vert \vert x_{id}^{{}^{tex}} \vert \vert} \ ) \\
x_{id}'^{{}^{ \ PCA}} & = \ \mathcal{A} \ ( \ x_{id}^{{}^{PCA}} \ )
\end{split}
\end{equation}

To construct the anonymized identity encoding $x_{id}'$ from $x_{id}'^{{}^{ \ PCA}}$, a weighted average of the nearest $j$ real samples, where similarity is measured in the lower dimension PCA space, is computed: 

\begin{equation}
x_{id}' = z_{id}[1 : j] \ * \ \sigma \left( \lambda \ d_{\angle} (x_{id}'^{{}^{\ PCA}}, z_{id_{\ j}}^{{}^{\ PCA}}) \right)
\end{equation}

In our experiments, parameters used were $j = \{1, \dots, 8\}$, $\sigma$ is the softmax equation, and $\lambda = 32$. 

This adaptation allows the resampling of identity despite the model's sparsely populated embedding space.  A limitation is the added requirement to maintain a set of representative embeddings in the original domain, but this adaptation enables the adaptations of our proposed privacy mechanisms to a broader set of generative architectures without retraining.

\subsection{Evaluation Pipeline}

\begin{figure}[t]
    \centering
    \includegraphics[width=1\linewidth]{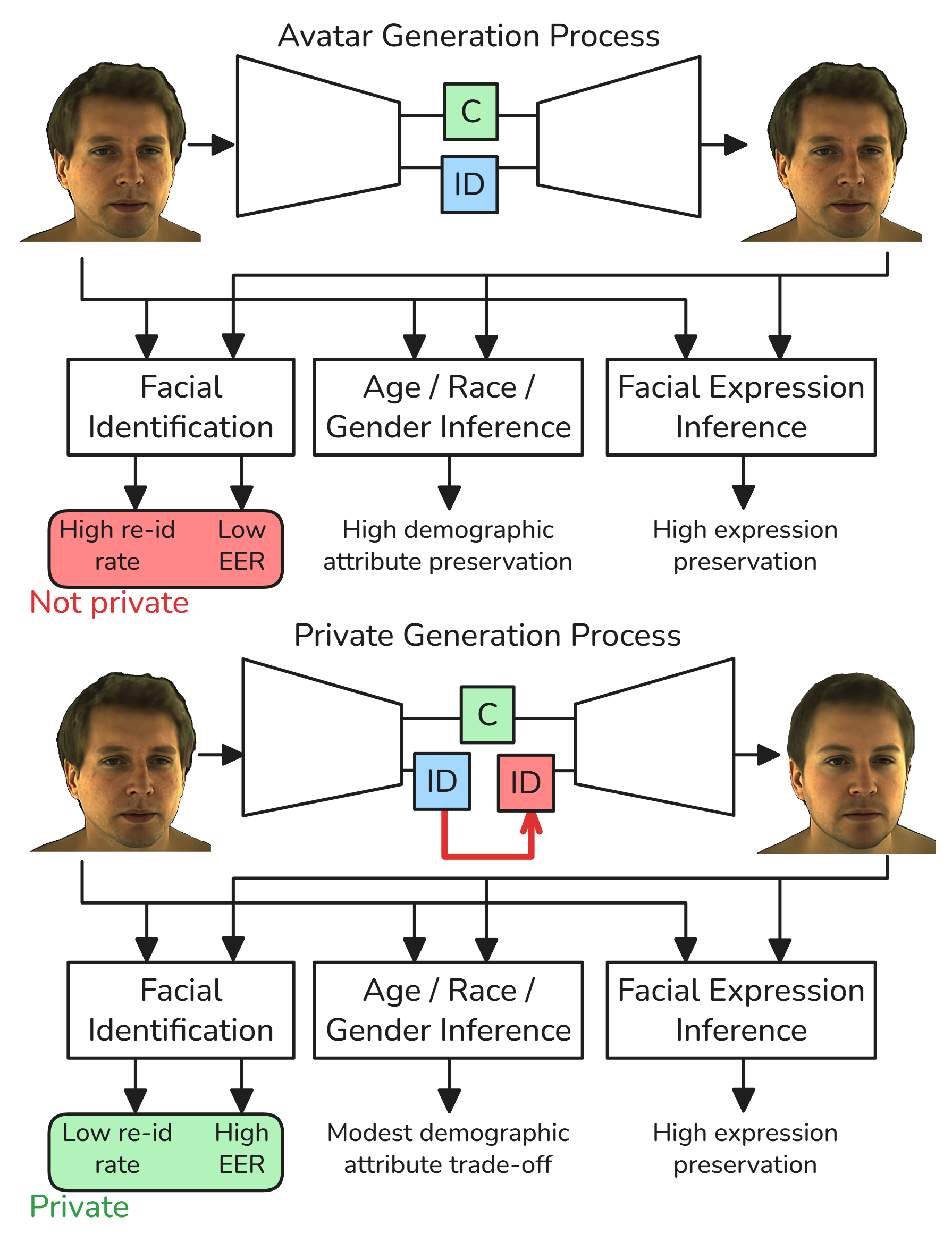}
    \caption{The evaluation pipeline utilized.  Facial identification and demographic / expression matching are evaluated by comparing generated appearances to source images.  Our approaches alter encoded identity features midway through the generation process to protect the identity of rendered avatars.  For 2D face synthesis, we evaluate on large-scale datasets and compare results against a number of baselines.  When evaluating 3D avatars, 2D renders of the avatars are used for evaluation.}
    \label{fig:evaluation_block_diagram}
\end{figure}

Our private avatar synthesis framework is evaluated across axes of privacy protection and utility preservation.  Utility is explored by measuring the impacts to information contained within visual appearance (the information encoded inside the identity embedding, such as demographic attributes), and attributes unrelated to visual appearance (such as expression preservation).  A high-level overview of the evaluation pipeline is shown in \autoref{fig:evaluation_block_diagram}.

To our knowledge, this approach is the first to explore privacy-centric rendering techniques for photorealistic 3D avatars.  Because of this, the bulk of our comparative evaluation is against established techniques for the privacy of 2D faces, particularly those rooted in differential privacy theory.  We additionally analyze our results on 3D avatars by constructing a small dataset of 3D avatars rendered to 2D face portraits, showing consistent privacy protections with our larger 2D evaluations.

\noindent\textbf{Privacy Evaluation.}

Privacy protections are evaluated by measuring two metrics common in face de-identification literature.  Rank $k$ identity matching compares a query face against a reference database of portrait images.  Results are sorted by $d_{\angle}$ between identity embeddings, storing \texttt{true} if the query's identity is sorted within $k$ samples, \texttt{false} otherwise.  Results are reported at $k=1$ and $k=50$, with the latter showing a more challenging benchmark for privacy performance.  To measure face verification, the equal error rate (EER) of face matching is computed by fitting a threshold to a collection of matching and unmatching face pairs.  We evaluate EER after de-identifying the query face against an unaltered enrollment face.  When measuring privacy, a lower rank $k$ metric and higher EER metric is preferred.

For rank $k$, the CelebA dataset~\cite{liu_deep_2015} test split containing $1000$ identities and $20k$ images is used.  EER is computed using the Labeled Faces in the Wild dataset~\cite{huang_labeled_2008} with $6000$ total face pairs (equally split between matching and unmatching pairs).  For 3D avatar evaluation, a face registry database is constructed by projecting avatar renders to 2D images.  One image is rendered per individual ($N = 256$) from a frontal view, with varied facial expressions across individuals.

\noindent\textbf{Utility Evaluation.}

One of a user's (and other users viewing their avatars') primary notions of utility is that the avatar remains high fidelity and natural.  The expressive content should be preserved so that user actions and intent can still be perceived.  Additionally, attributes core to the user's identity, such as their age, race, gender, or aesthetics, should be preserved~\cite{freeman_body_2021}.

Facial utility is measured with multiple face analysis algorithms\footnote{Gender, age, and race classifiers are trained using the VGGFace architecture~\cite{parkhi_deep_2015} and the emotion classifier is a small fully connected network.  All are implemented through the Deepface codebase: \url{https://github.com/serengil/deepface}~\cite{serengil_hyperextended_2021}.}, measuring the average differences between the true and privatized faces.  CelebA's test split and the same 3D avatar set are used for computing utility.  The following utility metrics are computed:

\begin{itemize}
    \item Emotion classification (anger, fear, neutral, sadness, disgust, happiness and surprise).
    \item Gender classification (male, female).
    \item Age prediction (0 - 100 years).
    \item Race classification (Asian, White, Middle Eastern, Indian, Latin and Black).
    \item Structural similarity index measure (SSIM), a common image quality metric computing the difference in contrast, structure, and luminance of the two images.
\end{itemize}

\noindent\textbf{Baselines.}

Multiple baselines, each providing differential privacy guarantees on privatized face images, are implemented to serve as comparison points.  PixelDP is a method which pixelizes an image then adds Laplace noise to each downsampled pixel~\cite{fan_image_2018}.  MetricSVD uses a metric privacy formulation to de-identify images with visual results similar to blurring~\cite{fan_practical_2019}. MetricSVD decomposes a face image using SVD, discards all but $k$ most significant values, then adds sampled noise before reconstructing the image.  PixelDP and MetricSVD are both constrained to image manipulations, and a strong and weak variant is evaluated for each.

The final baseline, IdentityDP, adds 1-D noise sampled from the Laplace distribution to the embedded identity features of a face synthesis model~\cite{wen_identitydp_2022}.  IdentityDP and \epsmethod{} both offset feature-space identity embeddings with a LDP formulation.  While IdentityDP works well using an Arcface backend model~\cite{deng_arcface_2019}, it fails to generalize when assumptions are violated (discussed in \autoref{sec:comp_identitydp}).

\section{Results}
\label{sec:results}

First, the privacy-preserving capabilities of \epsmethod{} and \thetamethod{} are validated.  Then, the level of facial utility maintained after anonymization is presented.  Image-based results are tabulated in \autoref{tab:2D_results} (SimSwap) and \autoref{tab:2D_results_ghost} (GHOST), and codec avatar-based results are tabulated in \autoref{tab:avatar_results}.

\begin{table*}[t]
\caption{Privacy / utility results of privacy operations on 2D face images using the SimSwap face swapping architecture with comparisons to IdentityDP, PixelDP, and MetricSVD.  Down arrows ($\downarrow$) indicate that a lower identification accuracy or utility metric is ideal, and up arrows ($\uparrow$) indicate that a higher EER, utility classification accuracy, or similarity metric is ideal.  \textbf{Random sampling} over $\mathbb{S}^{n-1}$ is a lower bound for \epsmethod{}, reflecting any identity leakage in the generative model, and \textbf{image reconstruction} represents the model's attempt to reconstruct each individual's appearance without modification.}
\label{tab:2D_results}
\begin{tabular}{lrrrrrrrr}
\toprule
Method & Rank 1 (\%) $\downarrow$ & Rank 50 (\%) $\downarrow$ & EER (\%) $\uparrow$ & SSIM $\uparrow$ & Age (diff.) $\downarrow$ & Race (\%) $\uparrow$ & Gender (\%) $\uparrow$ & Emotion (\%) $\uparrow$ \\
\midrule
\textbf{Real Images} & 98.13 & 98.52 & 2.95 & 1.00 & 0.00 & 100.0 & 100.0 & 100.0 \\
\textbf{Img. reconst.} & 84.50 & 92.92 & 4.67 & 0.92 & 3.45 & 78.57 & 91.05 & 74.10 \\ \midrule
\epsmethod{} $\varepsilon$=200 & 54.91 & 83.14 & 8.93 & 0.91 & 4.31 & 67.22 & 74.96 & 73.55 \\
\epsmethod{} $\varepsilon$=100 & 34.29 & 69.81 & 11.42 & 0.91 & 4.66 & 62.96 & 67.80 & 73.13 \\
\epsmethod{} $\varepsilon$=50 & 23.61 & 58.46 & 13.15 & 0.91 & 4.80 & 61.27 & 64.72 & 72.46 \\
\epsmethod{} $\varepsilon$=10 & 16.17 & 48.62 & 17.82 & 0.91 & 4.90 & 61.17 & 63.94 & 72.55 \\
\epsmethod{} $\varepsilon$=1 & 14.23 & 46.55 & 16.90 & 0.91 & 4.87 & 60.22 & 64.02 & 72.68 \\ \midrule
\textbf{Rand. sampling} & 14.54 & 46.78 & 17.57 & 0.91 & 4.87 & 60.52 & 64.20 & 72.76 \\ \midrule
\thetamethod{} $\theta$=60$\degree$ & 72.00 & 89.60 & 5.93 & 0.92 & 3.86 & 72.34 & 83.37 & 73.97 \\
\thetamethod{} $\theta$=90$\degree$ & 13.87 & 46.46 & 17.44 & 0.91 & 4.99 & 60.14 & 63.94 & 72.21 \\
\thetamethod{} $\theta$=135$\degree$ & 1.52 & 7.40 & 54.26 & 0.90 & 4.96 & 68.04 & 85.91 & 71.83 \\
\thetamethod{} $\theta$=150$\degree$ & 1.25 & 6.46 & 59.23 & 0.90 & 5.03 & 69.92 & 87.70 & 71.56 \\ \midrule
IdentityDP ($\varepsilon = 100$) & 23.98 & 59.90 & 13.80 & 0.91 & 4.81 & 61.28 & 65.13 & 72.02 \\
IdentityDP ($\varepsilon = 1$) & 14.85 & 46.38 & 16.94 & 0.91 & 5.03 & 60.29 & 63.97 & 72.45 \\
PixelDP (weak) & 21.75 & 55.53 & 29.04 & 0.81 & 7.96 & 53.05 & 80.57 & 26.72 \\
PixelDP (strong) & 0.28 & 7.01 & 46.47 & 0.74 & 6.34 & 59.42 & 50.41 & 19.64 \\
MetricSVD (weak) & 95.31 & 98.11 & 4.03 & 0.94 & 3.70 & 76.87 & 70.14 & 73.11 \\
MetricSVD (strong) & 9.25 & 27.18 & 37.62 & 0.83 & 6.55 & 51.64 & 54.22 & 27.00 \\
\bottomrule
\end{tabular}
\end{table*}

\begin{table*}[t]
\centering
\caption{Privacy / utility results of privacy operations applied to the Relightable Gaussian Codec Avatars architecture and evaluated on a dataset of rendered poses (256 identities / poses).}
\label{tab:avatar_results}
\begin{tabular}{lrrrrrr}
\toprule
Method & Rank 1 (\%) $\downarrow$ & SSIM $\uparrow$ & Age (diff.) $\downarrow$ & Race (\%) $\uparrow$ & Gender (\%) $\uparrow$ & Emotion (\%) $\uparrow$ \\
\midrule
\textbf{Img. reconst.} & 98.05 & 0.86 & 3.28 & 78.91 & 89.06 & 64.06 \\ \midrule
\epsmethod{} $\varepsilon$=50 & 87.50 & 0.83 & 3.96 & 73.05 & 84.38 & 59.77 \\
\epsmethod{} $\varepsilon$=10 & 20.31 & 0.76 & 4.69 & 58.20 & 70.31 & 50.39 \\
\epsmethod{} $\varepsilon$=1 & 3.91 & 0.72 & 5.08 & 51.17 & 65.23 & 46.09 \\ \midrule
\textbf{Rand. sampling} & 1.95 & 0.71 & 5.20 & 46.09 & 57.03 & 48.05 \\ \midrule
\thetamethod{} $\theta$=60$\degree$ & 16.02 & 0.76 & 5.06 & 55.86 & 70.31 & 48.44 \\
\thetamethod{} $\theta$=90$\degree$ & 1.95 & 0.70 & 5.45 & 37.89 & 58.59 & 45.31 \\
\thetamethod{} $\theta$=135$\degree$ & 2.34 & 0.68 & 5.41 & 36.72 & 46.09 & 40.62 \\
\thetamethod{} $\theta$=150$\degree$ & 0.78 & 0.67 & 5.58 & 34.77 & 41.41 & 39.06 \\
\bottomrule
\end{tabular}
\end{table*}

\begin{table*}[t]
\caption{Privacy / utility results of privacy operations on 2D face images using the GHOST face synthesis architecture.  This table follows the format of \autoref{tab:2D_results} and shows consistent trends for privacy mechanisms when applied to the GHOST or SimSwap architectures.}
\label{tab:2D_results_ghost}
\begin{tabular}{lrrrrrrrr}
\toprule
Method & Rank 1 (\%) $\downarrow$ & Rank 50 (\%) $\downarrow$ & EER (\%) $\uparrow$ & SSIM $\uparrow$ & Age (diff.) $\downarrow$ & Race (\%) $\uparrow$ & Gender (\%) $\uparrow$ & Emotion (\%) $\uparrow$ \\
\midrule
\textbf{Real Images} & 98.13 & 98.52 & 2.95 & 1.00 & 0.00 & 100.00 & 100.00 & 100.00 \\
\textbf{Img. reconst.} & 97.80 & 98.21 & 3.48 & 0.96 & 2.52 & 85.42 & 93.67 & 76.97 \\
\midrule
\epsmethod{} $\varepsilon$=100 & 50.66 & 83.86 & 7.48 & 0.94 & 4.74 & 58.34 & 55.05 & 71.74 \\
\epsmethod{} $\varepsilon$=10 & 17.53 & 50.18 & 15.43 & 0.93 & 5.01 & 55.75 & 52.54 & 70.75 \\
\epsmethod{} $\varepsilon$=1 & 15.37 & 46.15 & 16.11 & 0.93 & 4.93 & 56.12 & 52.34 & 70.67 \\
\midrule
\textbf{Rand. sampling} & 14.96 & 45.97 & 16.90 & 0.93 & 4.99 & 54.82 & 52.40 & 70.59 \\
\midrule
\thetamethod{} $\theta$=60$\degree$ & 93.79 & 97.78 & 3.61 & 0.95 & 3.81 & 70.24 & 71.14 & 73.92 \\
\thetamethod{} $\theta$=90$\degree$ & 14.64 & 45.88 & 16.68 & 0.93 & 5.02 & 55.57 & 51.97 & 70.65 \\
\thetamethod{} $\theta$=135$\degree$ & 0.03 & 0.47 & 79.39 & 0.93 & 4.72 & 68.19 & 80.80 & 70.57 \\
\thetamethod{} $\theta$=150$\degree$ & 0.04 & 0.43 & 85.83 & 0.93 & 4.67 & 71.50 & 87.53 & 70.63 \\
\bottomrule
\end{tabular}
\end{table*}

\subsection{Privacy Results}

\epsmethod{}, which samples new identity embeddings by drawing from the VMF distribution, has provable $\varepsilon d_{\angle}$ privacy and $\varepsilon$ LDP guarantees.  This is empirically validated by comparing to uniform random sampling of the identity vector, which models that any output identity embedding is equally likely to have been sourced from any input.  When $\varepsilon = 1$, the rank 1 accuracy is $14.23$\%, and EER is equal to $16.9$\%.  These values approach random sampling's $14.54$\% and $17.57$\% respective accuracies.  As $\varepsilon$ increases, values converge towards the model's image reconstruction performance as the average outputs become closer to the true model reconstruction without noise.

\begin{figure}[h]
    \centering
    \includegraphics[width=\linewidth]{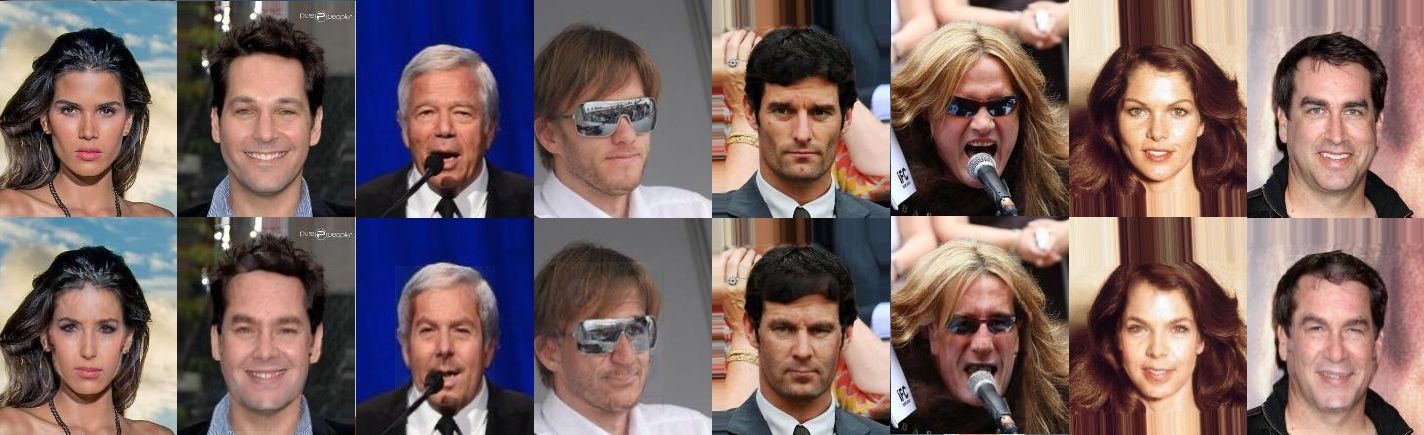}
    \caption{Failure cases (using the SimSwap backend).  Despite a fully random identity vector, imperfect identity disentanglement allows a facial recognition model to match faces before and after processing.}
    % Removing the portion about accessories to prevent confusion.  I don't concretely know if accessories are a prominent cause of failure, and that'd be a full experiment to prove.
    %Many failure cases are obscured with accessories, or the result has minimal changes in appearance.}
    \label{fig:failure_cases}
\end{figure}

For privacy mechanisms operating on identity embeddings, an assumption made of the generative model is that identity is well-disentangled.  This assumption is partially violated for both SimSwap and GHOST, evidenced by the rank 1 ID rate of random sampling being higher than random chance (random sampling equals $14.54\%$ for SimSwap, $14.96\%$ for GHOST).  A recent meta-analysis of Arcface identity encoders quantified noticeable identity leakage~\cite{shiohara_blendface_2023} and proposed innovations to mitigate leakage.  Both SimSwap and GHOST utilize Arcface models, so our evaluation inherits this flaw, but the higher floor for re-identification rate is not attributed to the proposed privacy algorithms.  A set of failure cases where randomly sampled identity embeddings failed to generate a de-identified result is visible in~\autoref{fig:failure_cases}.

\thetamethod{} offers greater control than \epsmethod{} for the level of identity offset per individual sample.  We see that $\theta = 90\degree$ is nearly equivalent to random sampling, with rank 1 = $13.87$\% and EER = $17.44$\%.  At $\theta > 90\degree$, there is an enforced dissimilarity, and identification accuracy continues to drop, reaching a rank 1 ID rate of $1.25$\% and EER of $59.23$\% at $\theta = 150\degree$.  Note that while $\theta = 180\degree$ would offer the greatest empirical performance, it is uniquely reversible by repeating the operation, as moving to the furthest possible point is a deterministic operation.

For 3D avatar generation, \epsmethod{} shows the same trends, but we see that the lower bound of our method by random sampling, $1.95$\%, is lower and nearer to actual random chance selection of $1 / 256$ (i.e., the codec avatar model does not suffer identity leakage).  Note that the $\varepsilon$ parameter in VMF is tied to the dimensionality of the projected hypersphere, so the embedding space in $\mathbb{R}^{16}$ rather than $\mathbb{R}^{512}$ changes the interpretation of $\varepsilon$ values.

\subsection{Utility Results}

For 2D face synthesis, there are negligible differences in utility for non-identity related features.  SSIM and emotion classification accuracy remain high, with negligible decreases compared to image reconstruction ($0.02$ max decrease for SSIM and $2.54$\% for emotion classification).  Age, race, and gender are visible appearance attributes, so by manipulating identity features, changes are expected to occur for these attributes.  For operations near uniform random sampling (\epsmethod{} $\varepsilon = 1$ and \thetamethod{} $\theta = 90\degree$), age difference $\simeq$ $4.87$ years, race classification $\simeq$ $60.52$\%, and gender classification $\simeq$ $64.20$\%.  While gender classification is somewhat close to random chance accuracy ($1 / 2$ for the implemented gender classifier), other attributes remain at levels much higher than chance.  By relaxing the privacy constraint by increasing $\varepsilon$ or lowering $\theta$, the amount of privacy granted can be balanced against the preservation of appearance attributes.

For 3D avatars, age, race, and gender follow the same trends as the 2D face evaluation.  Yet, there is a sharper decrease in emotion classification and SSIM.  Rendered 2D images of generated avatars were used for evaluation, but the domain is inherently different, possibly influencing the utility classifiers' performance.  While age, race, and gender models make inferences on larger images, the emotion classifier infers on downsampled greyscale images ($48 \times 48$)~\cite{serengil_hyperextended_2021}, so it could be particularly sensitive to the domain change.  SSIM also decreases as avatars become dissimilar; this can be attributed to the greater differences in skin / hair color for full avatar manipulations than when only manipulating the face region.

\subsection{Comparison to PixelDP and MetricSVD}

\begin{figure}[h]
    \centering
    \includegraphics[width=\linewidth]{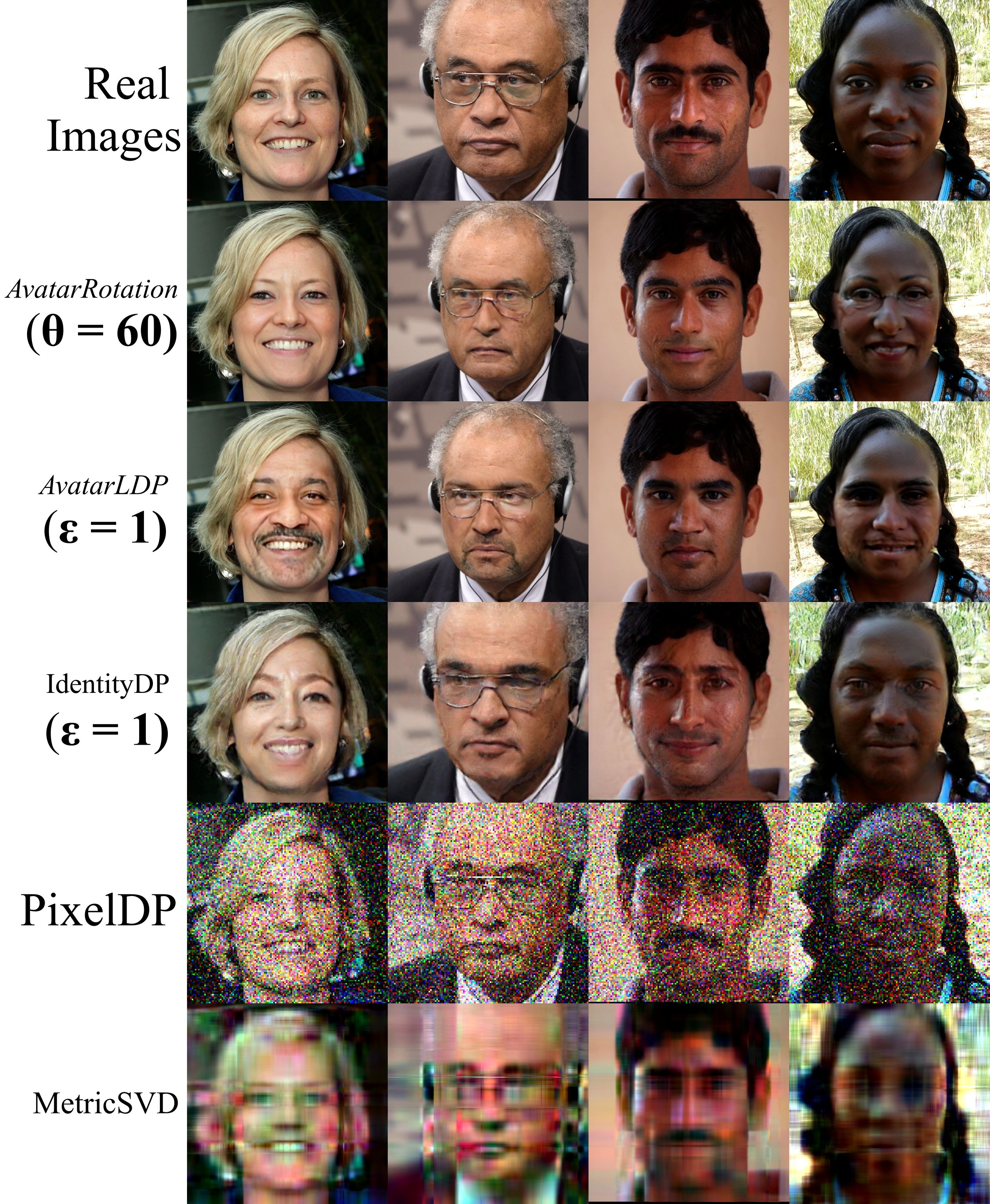}
    \caption{Visual comparison between our privacy mechanisms and baselines which provide differential privacy guarantees.  Medium-strength parameters for PixelDP and MetricSVD are rendered for visualization.}
    \label{fig:baseline_results}
\end{figure}

A visual comparison against baselines is seen in \autoref{fig:baseline_results}.  While PixelDP~\cite{fan_image_2018} and MetricSVD~\cite{fan_practical_2019} methods are able to protect privacy when implemented with strong parameters, they greatly degrade visual utility.  Additionally, these operations are based on point-in-time images, so could not be applied to video frames in a consistent manner. 

\subsection{Comparison with IdentityDP}
\label{sec:comp_identitydp}

\begin{figure}[h]
    \centering
    \includegraphics[width=\linewidth]{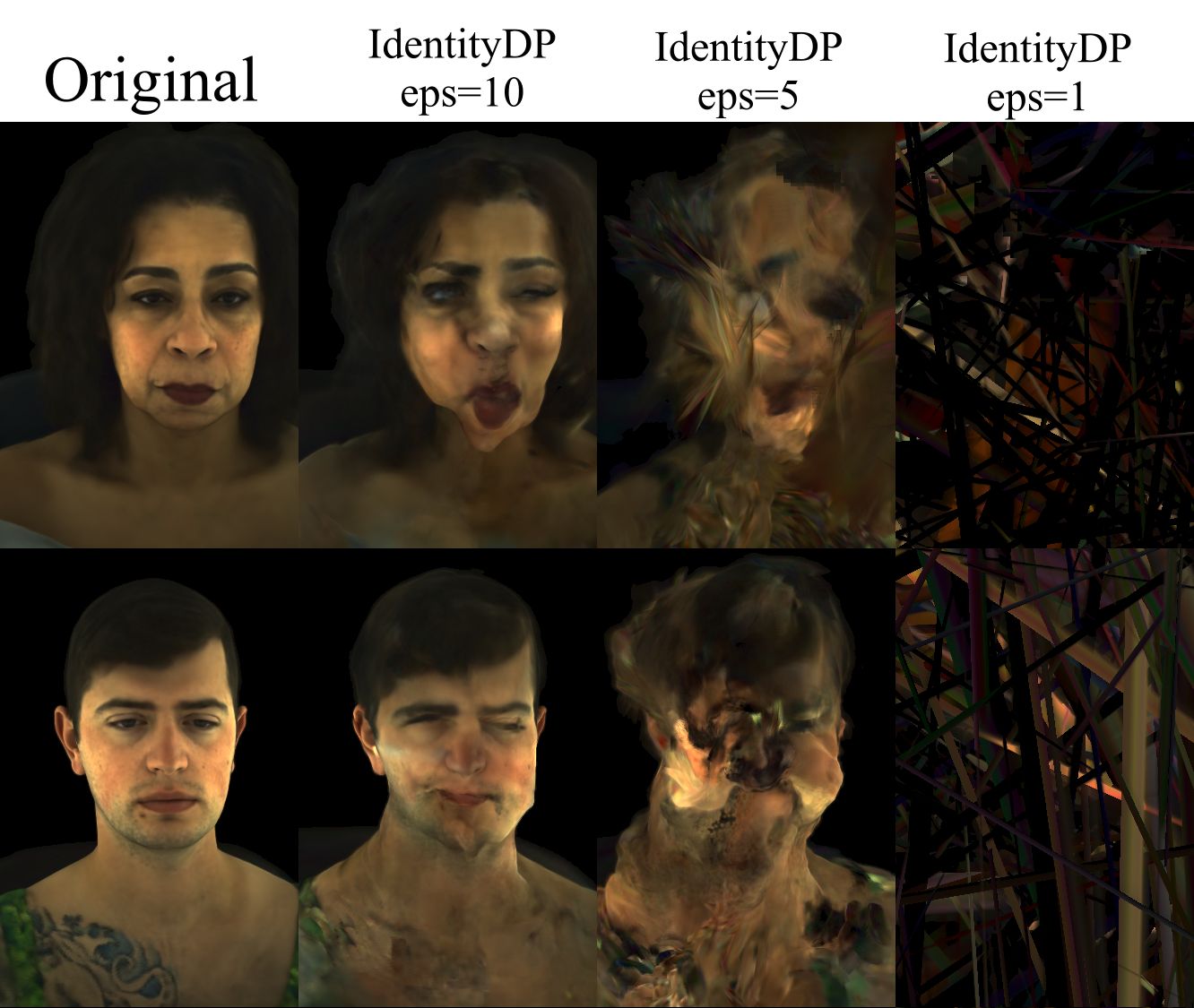}
    \caption{Results when applying IdentityDP to the codec avatar model.  Because the codec avatar model encodes identity into an arbitrary space, IdentityDP's additive noise pushes samples outside the model's known distribution, inducing model collapse.}
    \label{fig:identitydp_collapse}
\end{figure}

\epsmethod{} and IdentityDP operate similarly when applied to the identity embedding space of the SimSwap model.  Where IdentityDP's use of the Laplace mechanism adds independently-sampled 1-D noise to the entire feature vector, \epsmethod{}'s sampling from the VMF distribution disperses results evenly over the nearby surface of the hypersphere.  When applied inside a purely angular metric space (where magnitude is not considered as a feature), such as Arcface~\cite{deng_arcface_2019}, IdentityDP randomizes the direction of the vector through the independently sampled noise.  However, if applied to models that contain meaningful magnitude features, IdentityDP can create unintelligible output by resampling outside the plausible embedding region.  Attempts to apply IdentityDP to the codec avatar models resulted in model collapse, as seen in \autoref{fig:identitydp_collapse}.

\section{Discussion}
\label{sec:discussion}

In this work, we introduced a methodology to obfuscate user appearance during the generation of photorealistic self-avatars.  We developed two distinct privacy mechanisms which enable controlled manipulations encoded identity representations.  The composition of these two techniques would allow for even greater control, balancing differential privacy guarantees with precise offsets.  Our methods can protect user privacy while largely preserving broader attributes and photorealism.  

As photorealistic self-avatars become more prevalent in MR, there will be a need to protect user identities while delivering a consistent, photorealistic experience across a wide variety of applications.  To achieve this, we theorized sectioning online interactions into either \textit{secure} or \textit{public} interaction environments and explained the need for appearance protections in the public setting, where users may interface with unknown peers or less trustworthy applications.  These privacy operations can help to protect the privacy of users while preserving an overall sense of self and a photorealistic style in public interaction environments, and could greatly mitigate internet harms in social MR. 

Generating differentially private 3D self avatars (via \epsmethod{}) gives all appearances a level of uncertainty, preventing adversaries from making impersonation attempts on specific user likenesses or learning user appearances.  \thetamethod{} can ensure that each individual sample is different from the user's true appearance and could enable user-specified privacy controls where users offset their appearance to a preferred amount.  A simple composition of methods, \epsmethod{} with high $\varepsilon$ followed by \thetamethod{}, would inject uncertainty as to appearance while specifying the offset for the sampling distribution.  This composition preserves the LDP guarantee through its post-processing property~\cite{dwork_algorithmic_2014}.

\subsection{Real-world Implementations}

Choosing the appropriate parameters for a privacy mechanism can be a challenge in its own right, and is highly application dependent.  For differential privacy guarantees, finding the optimal choice of $\varepsilon$ for a given application is an active area of research~\cite{near_guidelines_2025}.  An important consideration is the fact that user re-identification rate is highly correlated with the number of potential identities to match~\cite{friedman_biometric_2022}.  In our evaluations, CelebA provided a reference set of $1000$ identities~\cite{liu_deep_2015}, and low $\varepsilon$ / high $\theta$ were necessary in order to greatly diminish re-identification rates.  Yet, if we envision a Metaverse ecosystem with millions of users, more modest privacy parameters can be chosen, improving the privacy-utility trade-off.  The parameters chosen would ultimately be determined by the mixed reality platforms that supply devices capable of encoding and decoding avatars.

An assumption for deploying these privacy mechanisms is that appearance will be altered whenever donning an avatar in MR.  This could be implemented in a number of ways.  A new, private appearance could be generated and shared every time a user joins an application, or perhaps the operation occurs once at avatar creation, or on regular intervals at a long time-scale.  Mapping a user to a single de-identified appearance has a potential security risk; once an adversary learns this mapping, they again become able to track the user and could share their findings with other adversaries.  Yet, releasing a new appearance at every login could diminish user experiences at times.  Say two users consistently meet through a shared activity; re-learning each others' avatar appearances repeatedly might hinder their ability to form a connection.  For these reasons, providers implementing the proposed privacy protections must be aware of potential trade-offs within implementation details.

\subsection{Identity Disentanglement}

When analyzing SimSwap~\cite{chen_simswap_2020} and GHOST~\cite{groshev_ghostnew_2022}, some identity leakage was seen in empirical evaluations.  Random sampling of identity yielded $\sim$15\% re-identification rates, despite strong visual appearance changes.  Shiohara et al. recently attributed identity leakage to flaws in Arcface's encoder~\cite{shiohara_blendface_2023} and developed an improved model.  Additionally, a number of recent works in the closely related topic of audio voice conversion have developed methods to better disentangle speaker identity~\cite{shamsabadi_differentially_2023, choi_dddm-vc_2024}.  In future work, these techniques shall be incorporated to develop backend face synthesis models with better disentanglement.  This strengthens models' ability to satisfy the assumptions of the developed privacy mechanisms, and would directly improve empirical privacy results.

\subsection{User Identity Obfuscation via Other Data Streams}

Facial identity is objectively the most identifiable attribute of a photorealistic self-avatar.  Yet, many driving signals can also be used to identify users to a degree~\cite{meng_-anonymizing_2024} and are in need of privacy solutions.  These modalities include body motions contained in the avatar's animation skeleton~\cite{nair_deep_2024}, eye movement data~\cite{wilson_privacy-preserving_2024}, and spoken audio~\cite{panariello_speaker_2024}.

The aggregation of many small data records has been shown to robustly identify users~\cite{narayanan_robust_2008}.  In a similar fashion, adversarial training over multiple MR signals has recently been shown to increase re-identification~\cite{jarin_popets_nodate}.  Thus, it will be important to protect identity across \textbf{each} modality in future systems~\cite{ibragimov_toward_2025}.

\subsection{Improving Private Avatar Generation}

A remapping of the Relightable Gaussian Codec Avatar~\cite{saito_relightable_2024} model's identity features was required for privatization and reconstruction of embeddings.  This is an effect of the generative model's architecture, which itself is constrained by the availability of training data ($N = 256$).  In this case, identities were embedded sparsely with a U-net structure~\cite{ronneberger_u-net_2015} to encode as much information about individual users as possible.  Yet, later architectures could leverage far greater training sets, enabling 3D avatar architectures to embed identity densely following large-scale classifier methodology~\cite{deng_arcface_2019}.  Thus, public identity encoders could be designed specifically with the integration of privacy mechanisms in mind.  For example, a recent work by Bai et al. encoded over 17,000 identities sourced from phone scanes~\cite{bai_universal_2024}.  As \textit{universal} encoders become available, it will become more straightforward to apply privacy mechanisms and to synthesize novel privacy-preserving appearances.

% \section*{Acknowledgments}
% This should be a simple paragraph before the References to thank those individuals and institutions who have supported your work on this article.

\bibliographystyle{IEEEtran}
\bibliography{bibliography}

\end{document}